\documentclass[prd,twocolumn,floatfix,preprintnumbers,nofootinbib,superscriptaddress]{revtex4-2}

\usepackage{ulem}
\usepackage{bm}
\usepackage{times}
\usepackage{amssymb,amsbsy,amsmath,amsfonts}
\usepackage{graphicx}
\usepackage{float}
\usepackage{color}
\usepackage{morefloats}
\usepackage{rotating}
\usepackage{srcltx}
\usepackage{slashed}
\usepackage{subfigure}
\usepackage{multirow}
\usepackage{verbatim}
\usepackage{hyperref}
\usepackage{tabularx}
\usepackage{braket}
\usepackage{diagbox}
\usepackage{appendix}
\usepackage{subfigure}
\usepackage{makecell}
\usepackage{booktabs}
\usepackage[section]{placeins}

\usepackage[outdir=./]{epstopdf}

\begin{document}

\title{Dynamically generated states from the $\eta K^*\bar{K}^*$, $\pi K^*\bar{K}^*$, and $K K^*\bar{K}^*$ systems within the fixed-center approximation}

\author{Qing-Hua Shen}~\email{shenqinghua@impcas.ac.cn}
 \affiliation{Institute of Modern Physics, Chinese Academy of Sciences, Lanzhou 730000, China}
 \affiliation{School of Nuclear Sciences and Technology, University of Chinese Academy of Sciences, Beijing 101408, China}
 \affiliation{School of Physical Science and Technology, Lanzhou University, Lanzhou 730000, China}

\author{Xu Zhang}~\email{zhangxu@itp.ac.cn}
 \affiliation{CAS Key Laboratory of Theoretical Physics, Institute of Theoretical Physics, Chinese Academy of Sciences, Beijing 100190, China}
 
\author{Xiang Liu}~\email{xiangliu@lzu.edu.cn}
 \affiliation{School of Physical Science and Technology, Lanzhou University, Lanzhou 730000, China}
  \affiliation{Lanzhou Center for Theoretical Physics, Key Laboratory of Theoretical Physics of Gansu Province, Lanzhou University, Lanzhou 730000, China}
  \affiliation{Key Laboratory of Quantum Theory and Applications of MoE, Lanzhou University, Lanzhou 730000, China}
\affiliation{MoE Frontiers Science Center for Rare Isotopes, Lanzhou University, Lanzhou 730000, China}
\affiliation{Research Center for Hadron and CSR Physics, Lanzhou University and Institute of Modern Physics of CAS, Lanzhou 730000, China}  
  
\author{Ju-Jun Xie}~\email{xiejujun@impcas.ac.cn}
\affiliation{Institute of Modern Physics, Chinese Academy of Sciences, Lanzhou 730000, China} 
\affiliation{School of Nuclear Sciences and Technology, University of Chinese Academy of Sciences, Beijing 101408, China} 
\affiliation{Southern Center for Nuclear-Science Theory (SCNT), Institute of Modern Physics, Chinese Academy of Sciences, Huizhou 516000, Guangdong Province, China}
 
\begin{abstract}

The three-body systems $\eta K^* \bar{K}^*$, $\pi K^* \bar{K}^*$, and $K K^* \bar{K}^*$ are further investigated within the framework of fixed-center approximation, where $K^* \bar{K}^*$ is treated as the fixed-center, corresponding to the possible scalar meson $a_0(1780)$ or the tensor meson $f_2'(1525)$. The interactions between $\eta$, $\pi$, $K$, and $K^*$ are taken from the chiral unitary approach. The resonance structures appear in the modulus squared of the three-body scattering amplitude and suggest that $\eta / \pi / K$-$(K^*\bar{K}^*)_{a_0(1780)/f'_2(1525)}$ hadron state can be formed. By scattering the $\eta$ meson on the fixed-center $(K^* \bar{K}^*)_{a_0(1780)}$, it is found that there is a distinct peak around 2100 MeV, as shown in the modulus squared of the three-body scattering amplitude, which can be associated with the meson $\pi(2070)$. For the scattering of the $\eta$ meson on the $(K^* \bar{K}^*)_{f_2'(1525)}$, a resonance structure around 1890 MeV is found and it can be associated with the $\eta_2(1870)$ meson. Other resonance structures are also found and can be associated with $\pi_2(1880)$ and $\eta(2010)$.

\end{abstract}

\date{\today}

\maketitle

\section{Introduction}

With abundant observations of new hadronic states since 2003~\cite{Liu:2013waa,Hosaka:2016pey,Chen:2016qju,Richard:2016eis,Lebed:2016hpi,Guo:2017jvc,Olsen:2017bmm,Liu:2019zoy,Brambilla:2019esw}, the search for exotic hadronic matter has become a frontier of particle physics, which is also an effective way to deepen our understanding of the non-perturbative behavior of the strong interaction. These novel phenomena have also stimulated extensive discussion of the interaction between hadrons. A typical example is that the characteristic mass spectrum of $P_c$ states in the  $\Lambda^0_b \to K^- J/\psi p$ process~\cite{LHCb:2015yax} supports the hidden-charm molecular baryons composed of anti-charmed meson and charmed baryon~\cite{Wu:2010jy,Wu:2010vk,Wang:2011rga,Yang:2011wz}. Thus, it is also the reason why hadronic molecular state is popular for deciphering the nature of these new hadronic states.

In addition to the heavy flavor sector, recently there has been many studies of light flavor sector in the picture of the hadronic molecular state, typically including the development of the treatment of hadron-hadron scattering by the chiral effective Lagrangians combined with non-perturbative unitary techniques in coupled channels~\cite{Oset:2016lyh,Dong:2021juy,Dong:2021bvy}. 
In Refs.~\cite{Molina:2008jw,Geng:2008gx}, the vector-vector interactions are investigated within chiral unitary approach. The scalar mesons $f_0(1370)$ and $f_0(1710)$ and tensor mesons $f_2(1270)$, $f'_2(1525)$ and $K^*_2(1430)$ were considered as dynamically generated states~\cite{Molina:2008jw,Geng:2008gx}. Within the vector-vector molecular picture, the scalar meson $f_0(1710)$ and the tensor meson $f'_2(1525)$ couple mostly to the $K^*\bar{K^*}$ channel, and most of their properties can be well
explained~\cite{Geng:2008gx,Du:2018gyn,Nagahiro:2008bn,Branz:2009cv,Geng:2009gb,Geng:2010kma,MartinezTorres:2012du,Xie:2014gla,Molina:2019wjj,Wang:2021jub}. Additionally, in Refs.~\cite{Geng:2008gx,Du:2018gyn}, an isovector partner of the scalar meson $f_0(1710)$ is also predicted, with its mass around $1780$ MeV, which is hereafter refereed to as $a_0(1780)$. The $a_0(1780)$ state is also strongly coupled to the $K^* \bar{K}^*$ channel\footnote{Similar conclusions are found in Ref.~\cite{Wang:2022pin}, where these pseudoscalar-pseudoscalar coupled channels were considered. The mass of $a_0(1780)$ obtained in Ref.~\cite{Wang:2022pin} is smaller than that predicted in Refs.~\cite{Geng:2008gx,Du:2018gyn}. The scalar meson $a_0(1780)$ has recently been observed by experiments~\cite{BaBar:2021fkz,BESIII:2022npc}. In view of the $K^*\bar{K}^*$ molecular state, the production of scalar mesons $a_0(1780)$ in $D^+_s \to \pi^+ K^0_S K^0_S$ and $D^+_s \to \pi^0 K^+ K^0_S$ was theoretically studied in Refs.~\cite{Dai:2021owu,Zhu:2022wzk,Zhu:2022guw}, where the experimental measurements on the invariant mass distributions of the final states can be reproduced well.}.
In fact, the investigation is not limited to the two-hadron systems as presented in Refs. \cite{MartinezTorres:2011vh,MartinezTorres:2011gjk,Liang:2013yta,Zhang:2016bmy,Bayar:2013bta,MartinezTorres:2008gy,MartinezTorres:2009xb}, which are involved in these reported mesonic states, such as $\pi(1300)$, $K(1460)$, $\eta(1475)$, $\pi_1(1600)$, $\rho(1700)$, $\phi(2170)$, respectively. Here, light flavor three-meson systems were explored. 

In the research field of the few-body problem, the treatment of the three-body system continues to attract the attention of theorists. Indeed, there is growing evidence that some existing and newly observed hadronic states could be interpreted in terms of resonances or bound states of three hadrons~\cite{MartinezTorres:2020hus,Wu:2022ftm,Malabarba:2021taj,Debastiani:2017vhv,Malabarba:2022pdo}, and some of new hadronic states~\cite{Luo:2021ggs,Luo:2022cun,Ikeno:2022jbb,Wu:2020job,Wei:2022jgc,Ikeno:2022jbb,Bayar:2023itf} have also been predicted in three-body systems. Technical developments have been made in recent years, where various approaches including Gassian expansion method~\cite{Hiyama:2003cu}, solving Faddeev equations in the coupled channel approach~\cite{MartinezTorres:2007sr,MartinezTorres:2009cw,Khemchandani:2009aj}, and the fixed center approximation (FCA) have been proposed. The FCA has been employed before, in particular in the study of the $\bar{K}d$ interaction at low energies~\cite{Chand:1962ec,Barrett:1999cw,Deloff:1999gc,Kamalov:2000iy}. This approach is also used to study these multi-$\rho$ states~\cite{Roca:2010tf} and $K^*$-multi-$\rho$ states~\cite{Yamagata-Sekihara:2010muv}. Using the FCA in the $\Delta \rho \pi$ system, an interesting explanation of the $\Delta_{5/2^+}$ puzzle was proposed~\cite{Xie:2011uw}.

In Ref.~\cite{Shen:2022etd}, the light flavor three-body $\eta K^* \bar{K}^*$, $\pi K^* \bar{K}^*$, and $K K^* \bar{K}^*$ systems are partly investigated within the framework of FCA, where the fixed-center $K^* \bar{K}^*$ is treated as the scalar meson $f_0(1710)$. In fact, there exist other allowed combinations of the $K^* \bar{K}^*$ system. As in Ref.~\cite{Geng:2008gx}, an $h_1$ state with mass around $1800$ MeV and an $a_2$ state with mass about $1570$ MeV were also found, and they couple strongly to the $K^*\bar{K}^*$ channel. However, these two dynamically generated states from the vector meson-vector meson interaction cannot be clearly identified with any of the $h_1$ and $a_2$ states listed in the \textit{Review of particle physics} (RPP)~\cite{ParticleDataGroup:2022pth}. In the present work, we further investigate $\eta K^*\bar{K}^*$, $\pi K^*\bar{K}^*$ and $KK^*\bar{K}^*$ three-body systems by the FCA, and we take the scalar meson $a_0(1780)$ and tensor meson $f_2'(1525)$ as $K^*\bar{K}^*$ molecular states~\cite{Dai:2013uua,Dai:2015cwa,Xie:2015isa,Xie:2016ghe,Oset:2023hyt}, and then scatter the $\eta$, $\pi$ and $K$ mesons on the fixed-center of $K^*$ and $\bar{K^*}$. Since we consider only the $s$-wave interaction, the above three-body systems can have quantum numbers $J^{PC} = 0^{-+}$ or $2^{-+}$, which indicates that we will investigate the pseudoscalar and pseudotensor low-lying excited states in the $\eta K^*\bar{K}^*$, $\pi K^*\bar{K}^*$ and $KK^*\bar{K}^*$ three-body systems. Besides, these states with exotic quantum numbers for the $K(K^*\bar{K}^*)_{a_0(1780)}$ with the total isospin $I=\frac{3}{2}$ sector are also investigated. For the two-body scattering, we take the interactions between pseudoscalar mesons and vector mesons as obtained with the chiral unitary approach~\cite{Lutz:2003fm,Roca:2005nm}.

To end this introduction, we would like to mention that the fixed center approximation is an effective and practical way to study three-body systems, widely accepted in the literatures~\cite{Ren:2018pcd,Dias:2017miz,Bayar:2011qj,Xiao:2011rc,Durkaya:2015wra,Yamagata-Sekihara:2010muv,MartinezTorres:2018zbl,SanchezSanchez:2017xtl,Xie:2010ig,Xie:2011uw}. It assumes the existence of a bound state of two particles that interact strongly with each other, and that the wave function of the bound state is not significantly changed by the interaction of an outside particle with it. This occurs when the third particle is lighter than the two particles in the bound state, or when the third particle has low energy, causing it to have minimal impact on the wave function of the bound state. Moreover, the FCA is technically much simpler than performing full calculations of the Faddeev equations. Therefore, it is important to utilize the FCA in various three-body systems, especially in the light flavor sector, where rich and available experimental data exists. However, it is important to note that limitations of using the FCA in three-meson studies were discussed in Ref.~\cite{MartinezTorres:2010ax}.

This paper is organized as follows. In Sec. \ref{formalism}, we present the FCA method to the three-body $\eta K^*\bar{K}^*$, $\pi K^*\bar{K}^*$ and $K K^*\bar{K}^*$ systems. In Sec.~\ref{numericalresults}, the numerical theoretical results and discussions are presented. Finally, a short summary is followed.

\section{Formalism and ingredients}~\label{formalism}

We are interested in the three-body systems: $\eta K^*\bar{K}^*$, $\pi K^*\bar{K}^*$, and $KK^*\bar{K}^*$. To study the dynamics of these three-body systems, we obtain the three-body scattering amplitudes using the fixed-center approximation method. A basic feature of the FCA is that one has a fixed-center bound of two particles and one allows the multiple scattering of the third particle with this bound state, which should not be changed by the interaction of the third particle. In addition, the interaction of a particle with a bound state of a pair of particles at very low energies or below the threshold can be studied efficiently and accurately by means of the FCA for the three-particle system~\cite{Chand:1962ec,Barrett:1999cw,Kamalov:2000iy}. In the present work, we extend this formalism to include states above three-body mass threshold and apply it to $\pi (K^*\bar{K}^*)_{a_0(1780)/f'_2(1525)}$ systems. In this section, we summarize the deduction of the three-body scattering amplitudes in the FCA framework.

\subsection{Fixed-center approximation to the three-body system}

\begin{figure*}[htbp]
   \centering
   \includegraphics[scale=0.5]{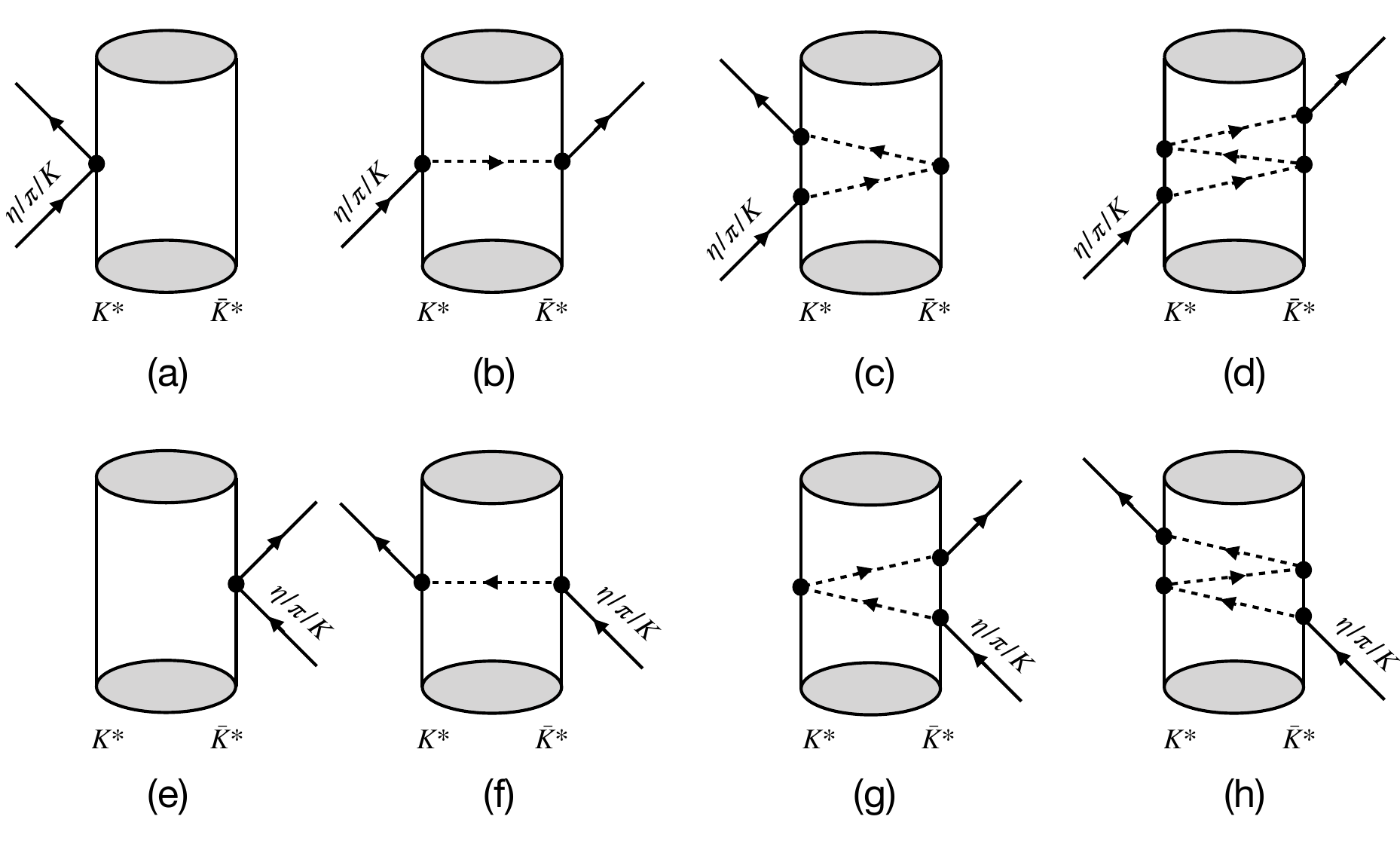}
   \caption{Schematic representation of the FCA for the Faddeev equations for the three-body sytstems $\eta K^*\bar{K}^*$, $\pi K^*\bar{K}^*$, and $KK^*\bar{K}^*$, with $K^*\bar{K}^*$ the fixed-center. In (a)–(d), the external particle $\eta/\pi/K$ initially colliders with the $K^*$, while in (e)–(h), the external particle $\eta/\pi/K$ initially colliders with the $\bar{K}^*$.}
   \label{fig:Feynman}
\end{figure*}

We will use the fixed-center approximation formalism to study the three-mesons system $\eta K^*\bar{K}^*$, $\pi K^*\bar{K}^*$, and $KK^*\bar{K}^*$, where we consider the $K^*\bar{K}^*$ as the fixed center, and treat it as a $a_0(1780)$ or $f_2'(1525)$ state. Then the $\eta$, $\pi$ or $K$ meson interacts with it. The corresponding diagrams are shown in Fig.~\ref{fig:Feynman}. In the following, we will refer to $K^*$, $\bar{K}^*$ and $\eta$ ($\pi$ or $K$) as particles 1, 2 and 3 respectively.

Following the formalism of Ref.~\cite{Roca:2010tf}, the three-body scattering amplitude $T$ for $\eta$ ($\pi$ or $K$) collisions with the fixed-center $K^* \bar{K}^*$ can be obtained by the sum of the partition functions $T_1$ and $T_2$:
\begin{eqnarray}
   T_1 &=& t_1+t_1G_0T_2, \\
   T_2 &=& t_2+t_2G_0T_1, \\
   T &=& T_1+T_2=\frac{t_1+t_2+2t_1t_2G_0}{1-t_1t_2G_0^2},  \label{eq:FCA}
\end{eqnarray}
with $T_1$ ($T_2$) the sum of all the diagrams in Fig.~\ref{fig:Feynman}, where the particle 3 collides firstly with the $K^*$ ($\bar{K}^*$) in the fixed-center. The $t_1(t_2$) represents the unitary two-body scattering amplitudes in coupled channels for the interactions of particle 3 with $K^*(\bar{K}^*)$.

In addition, $G_0$ is the loop function for particle 3 propagating between the $K^*$ and $\bar{K}^*$ inside the bound state ($B^*$), which can be written as
\begin{eqnarray}
G_0 = \frac{1}{2M_{B^*}} \int \frac{d^3 \vec{q}}{(2\pi)^3}\frac{F_{B^*}(q)}{(q^0)^2-|\vec{q}|^2-m_3^2+i\epsilon },
\end{eqnarray}
where $M_{B^*}$ and $F_{B^*}(q)$ are the mass and the form factor of the $K^*\bar{K}^*$ bound state, respectively. The $K^*\bar{K}^*$ bound state is treated in this paper as the scalar meson $a_0(1780)$ or the tensor meson $f_2'(1525)$, so $M_{B^*}$ is the mass of $a_0(1780)$ or $f'_2(1525)$. In addition, $m_3$ is the mass of the $\pi/\eta/ K$ meson. The $q^0$ is the energy of particle 3 with mass $m_3$ in the center-of-mass frame of particle 3 and $K^*\bar{K}^*$ bound state, which is given by
\begin{eqnarray}
q^0 = \frac{s+m_3^2-M_{B^*}^2}{2\sqrt{s}},
\end{eqnarray}
\noindent
where $s$ is the invariant mass square of the whole three-body system.

One of the ingredients in the calculation is the form factor for the assumed two-body $K^*\bar{K}^*$ bound state, the scalar meson $a_0(1780)$ and the tensor meson $f'_2(1525)$. Following the procedures as in Refs.~\cite{Roca:2010tf,Gamermann:2009uq,Yamagata-Sekihara:2010kpd}, one can obtain the expression of the form factor $F_{B^*}$ for the $s$-wave $K^*\bar{K}^*$ bound state $a_0(1780)$ or $f_2'(1525)$ as
\begin{eqnarray}
F_{B^*}(q) &=& \frac{1}{N}\int_{|\vec{p}|\leq \Lambda,|\vec{p}-\vec{q}|\leq\Lambda}d^3\vec{p}\frac{1}{2\omega_1(\vec{p})}\frac{1}{2\omega_2(\vec{p})}  \nonumber \\
    &&  \times\frac{1}{M_{B^*}-\omega_1(\vec{p})-\omega_2(\vec{p})} \frac{1}{2\omega_1(\vec{p}-\vec{q})}\frac{1}{2\omega_2(\vec{p}-\vec{q})}  \nonumber \\
     && \times\frac{1}{M_{B^*}-\omega_1(\vec{p}-\vec{q})-\omega_2(\vec{p}-\vec{q})},
\end{eqnarray}
where $\omega_1(\vec{p})=\omega_2(\vec{p})=\sqrt{|\vec{p}|^2+m_{K^*}^2}$, and the normalization factor $N$ is given by 
\begin{align}
   N=\int_{|\vec{p}| \leq \Lambda} d^3\vec{p} \left (\frac{1}{4\omega_1^2(\vec{q})}\frac{1}{M_{B^*}-2\omega_1(\vec{p})} \right ) ^2 .
\end{align}
The cutoff parameter $\Lambda$ is used to regularize the vector meson-vector meson loop functions in the chiral unitary approach~\cite{Geng:2008gx,Geng:2009gb}. In this work, the upper integration limit of $\Lambda$ has the same value as the cutoff used in Refs.~\cite{Geng:2008gx,Geng:2009gb}, with which one can obtain the scalar meson $a_0(1780)$ or the tensor meson $f_2'(1525)$ state in the vector meson-vector meson interactions in coupled channels.

The important ingredients in the calculation of the total scattering amplitude for the $\eta K^*\bar{K}^*$, $\pi K^*\bar{K}^*$, and $KK^*\bar{K}^*$ systems using the FCA are the two-body $\pi/\eta/K$-$K^*$, and $K^*\bar{K}^*$ unitarized $s$-wave interactions from the chiral unitary approach. Although the form of these interactions has been detailed elsewhere, we will briefly revisit them below for the case of $K^* \bar{K}^*$. This will allow us to review the general procedure for calculating the two-body amplitudes entering the FCA equations.

\begin{figure}[htbp]
   \centering
   \includegraphics[scale=0.42]{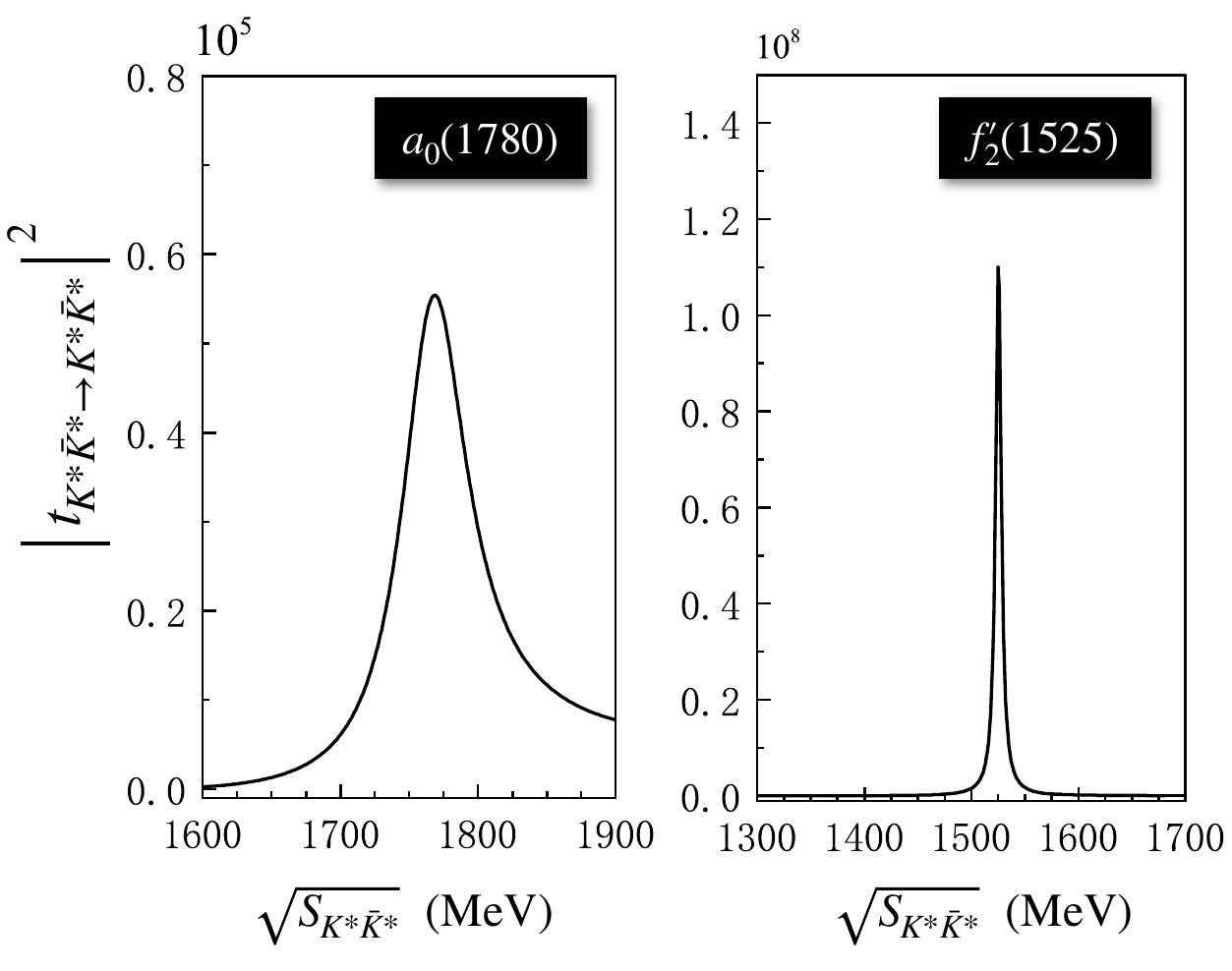}
   \caption{Modulus square of the $K^*\bar{K}^*\to K^*\bar{K}^*$ transition amplitude $t_{K^*\bar{K}^*\to K^*\bar{K}^*}$ in isospin=1 and spin=0 sector [$a_0(1780)$] and isospin=0 and spin=2 sector [$f'_2(1525)$] as a function of the invariant mass of the $K^*\bar{K}^*$ system.} 
   \label{fig:Tvv} 
\end{figure}

In Fig.~\ref{fig:Tvv}, the module squared of the transition amplitude $|t_{K^*\bar{K}^*\to K^*\bar{K}^*}|^2$ obtained from the chiral unitary approach in the coupled channels [$\rho \rho$, $\rho \omega$, $\rho \phi$, $K^*\bar{K}^*$ for Fig.~\ref{fig:Tvv} (a) and $K^*\bar{K}^*$, $\rho \rho$, $\omega \omega$, $\omega \phi$, $\phi \phi$ for Fig.~\ref{fig:Tvv} (b)] are shown. In these calculations, we use the cutoff regularization for the two-body vector meson-vector meson loop functions of $G_{VV}$, and the width of the vector meson $\rho$ and $K^*$ are taken into account. Furthermore, we take the same cutoff parameter $\Lambda = 1100$ (1009) MeV for all the channels in isospin=1 and spin=0 (isospin=0 and spin=2) sector. In addition, the obtained masses are 1769 and 1517 MeV for $a_0(1780)$ and $f'_2(1525)$, respectively, which are consistent with their masses quoted in the RPP~\cite{ParticleDataGroup:2022pth}.

Fig.~\ref{fig:Tvv} (a) shows the results obtained in the isospin=1 and spin=0 sector with cut off parameter $\Lambda=1100$ MeV, and Fig.~\ref{fig:Tvv} (b) shows the results obtained in the isospin=0 and spin=2 sector with $\Lambda = 1009$ MeV. From Fig.~\ref{fig:Tvv} (a), one can see a bump structure around 1780 MeV that can be assigned to the $a_0(1780)$ state. As discussed in the introduction part, the scalar meson $a_0(1780)$ was first observed by the BaBar Collaboration~\cite{BaBar:2021fkz} in 2021 and recently confirmed by the BESIII Collaboration~\cite{BESIII:2022npc}. 

It can be seen that, in Fig.~\ref{fig:Tvv} (b), the narrow peak around $1517$ MeV can be associated with the tensor meson $f_2'(1525)$. Comparing with the numerical results shown in Fig.~\ref{fig:Tvv} (a), it is found that the strength of $|t_{K^*\bar{K}^*\to K^*\bar{K}^*}|^2$ for $f'_2(1525)$ is much larger than that for the case of $a_0(1780)$. This indicates that the $s$-wave $K^*\bar{K}^*$ interaction in the isospin=0 and spin=2 sector is much stronger than that for the case of isospin=1 and spin=0. Furthermore, comparing the line shapes in Fig.~\ref{fig:Tvv} (a) with Fig.~\ref{fig:Tvv} (b), it is found that the width obtained for the tensor meson $f'_2(1525)$ is much narrower than that for the scalar meson $a_0(1780)$.

\begin{figure}[htbp]
   \centering
   \includegraphics[scale=0.32]{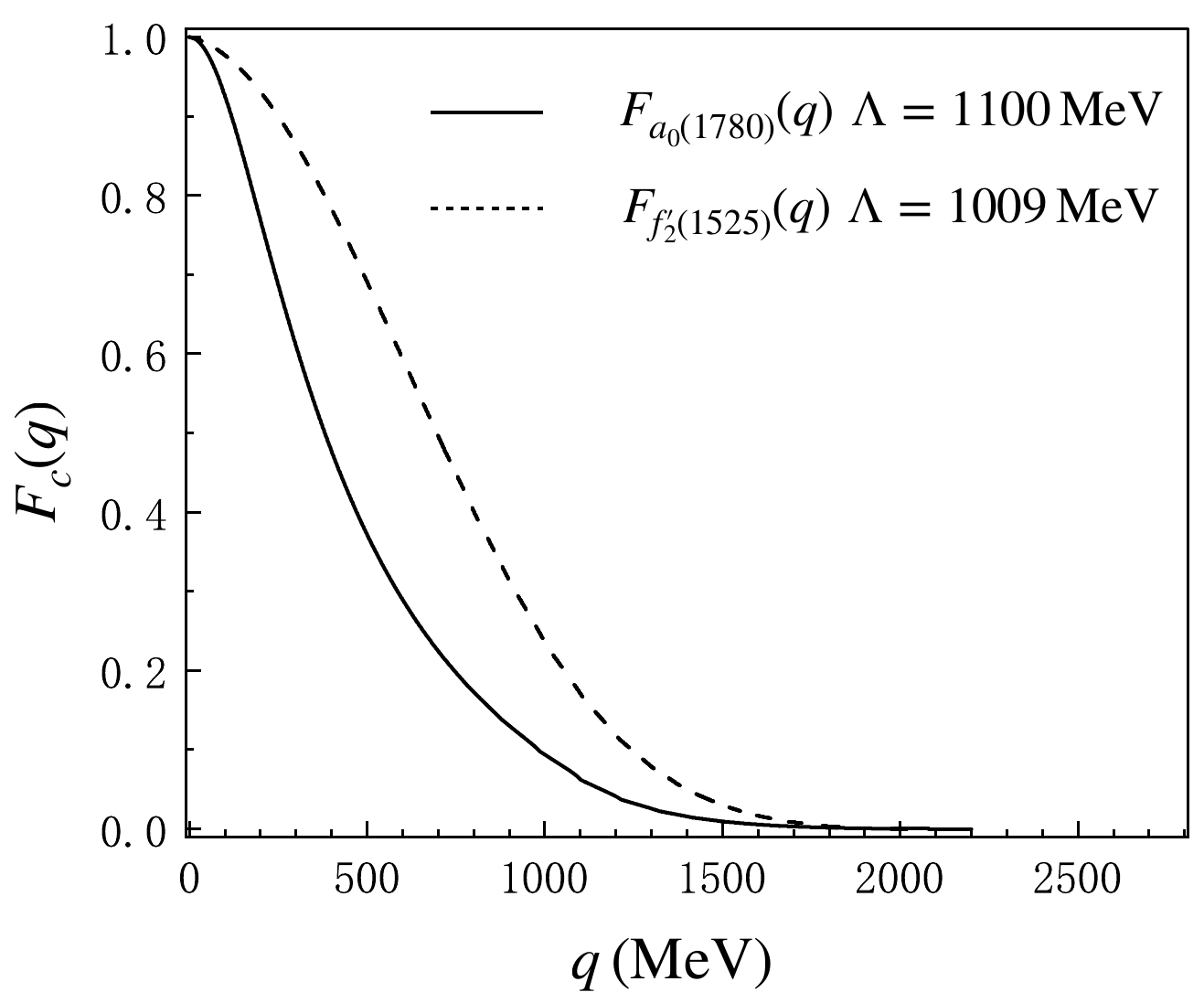}
   \caption{Theoretical results for the form factors of $(K^*\bar{K}^*)_{a_0(1780)}$ with $\Lambda=1100$ MeV (solid curve) and $(K^*\bar{K^*})_{f_2'(1525)}$ with $\Lambda=1009$ MeV (dashed curve).}
   \label{fig:F}
\end{figure}

Next, in Fig.~\ref{fig:F}, we show the numerical results for the respective form factors of $a_0(1780)$ (solid curve) and  $f_2'(1525)$ (dashed curve) as a function of $q=|\vec{q}|$, where the theoretical results are obtained for the $a_0(1780)[f_2'(1525)]$ with $\Lambda=1100(1009)$ MeV. The condition $|\vec{p} - \vec{q}| < \Lambda$ implies that the form factor $F_{B^*}(q)$ is exactly zero for $q > 2 \Lambda$. As discussed earlier, with these values of the cutoff parameter $\Lambda$, one can obtain $a_0(1780)$ and $f_2'(1525)$ resonances in vector mesons-vector meson coupled channel interactions as in Refs.~\cite{Geng:2008gx,Geng:2009gb}.

\begin{figure*}[htbp]
   \centering
   \includegraphics[scale=0.45]{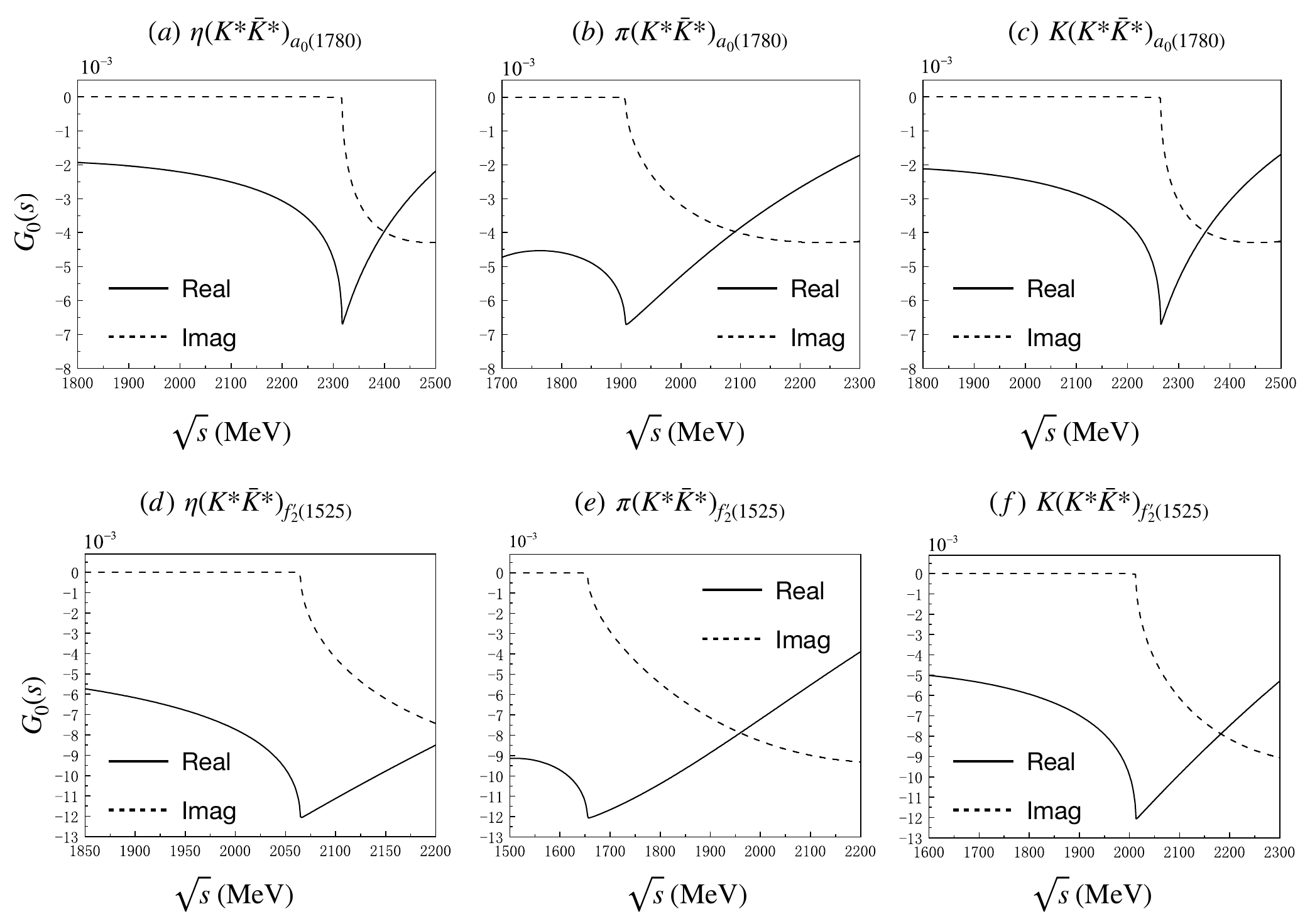}
   \caption{Real (solid line) and imaginary parts (dashed line) of the loop function $G_0$ for the $\eta/\pi/K(K^*\bar{K}^*)_{a_0(1780)}$ system with $\Lambda$=1100 MeV and for the $\eta/\pi/K(K^*\bar{K}^*)_{f_2'(1525)}$ system with $\Lambda$=1009 MeV.}
   \label{fig:G}
\end{figure*}

With the obtained form factors of the bound states $a_0(1780)$ and $f_2'(1525)$, then one can easily obtain the three-body loop function $G_0$ for the $\eta$ ($\pi$ and $K$) propagator between the $K^*$ and $\bar{K}^*$ of the bound states $a_0(1780)$ and $f_2'(1525)$, respectively. The $G_0$ depends on the invariant mass $\sqrt{s}$ of the whole three-body system. The real (Real) and imaginary (Imag) parts of the loop function $G_0$ are shown in Fig.~\ref{fig:G}. In this work, there are six $G_0$ functions and they are shown in Fig.~~\ref{fig:G} (a)-(f). Below the three-body mass threshold, the imaginary part of $G_0$ is zero. 

\subsection{Scattering of the $\eta$, $\pi$ and $K$ mesons on the $K^*\bar{K}^*$ system within the FCA}

According to Fig.~\ref{fig:Feynman} (a) and  (e), the single-scattering contributions of $t_1$ and $t_2$ are the appropriate combination of the two-body unitarized scattering amplitudes. For example, let us first consider the $\eta K^*\bar{K}^*$ system with the fixed-center $K^*\bar{K}^*$ as the $a_0(1780)$ state (denoted by $a_0$):
\begin{equation}
   \ket{10}_{\eta (K^*\bar{K}^*)_{a_0}}=\ket{0}_{\eta} \otimes \ket{10}_{a_0}
\end{equation}
\noindent
with
\begin{equation}
   \ket{10}_{a_0}=\frac{1}{\sqrt{2}}\left( \ket{\frac{1}{2},-\frac{1}{2}}+\ket{-\frac{1}{2},\frac{1}{2}}\right),
\end{equation}
\noindent
where the kets on the right side of the above equation represent $\ket{I_z^{K^*},I_z^{\bar{K}^*}}$ for $K^*\bar{K}^*$. Then the single-scattering contributions to the total amplitude of $\bra{\eta(K^*\bar{K}^*)_{a_0}} \hat{t}\ket{\eta(K^*\bar{K}^*)_{a_0}}$ can be easily obtained in terms of the unitary two-body transition amplitudes $t_{\eta K^* \to \eta K^*}$ and $t_{\eta \bar{K}^* \to \eta \bar{K}^*}$:
\begin{equation}
   \begin{aligned}
      &\bra{\eta(K^*\bar{K}^*)_{a_0}} \hat{t}\ket{\eta(K^*\bar{K}^*)_{a_0}}\\
     &=(\bra{A_1}+\bra{A_2})(\hat{t}_1+\hat{t}_2)(\ket{A_1}+\ket{A_2})\\
     &=\bra{A_1}\hat{t}_1\ket{A_1}+\bra{A_2}\hat{t}_2\ket{A_2}.
   \end{aligned}
\end{equation}
\noindent
Here, $\ket{A_1}$ stands for the state combined with $\eta$ and $K^*$, while $\ket{A_2}$ is the state of $\eta$ and $\bar{K}^*$. They are given by 
\begin{equation}
   \begin{aligned}
      \ket{A_1}&=\frac{1}{\sqrt{2}}\left( \ket{\frac{1}{2}\frac{1}{2},-\frac{1}{2}} +\ket{\frac{1}{2}-\frac{1}{2},\frac{1}{2}}\right), \ \\
      \ket{A_2}&=\frac{1}{\sqrt{2}}\left( \ket{\frac{1}{2}-\frac{1}{2},\frac{1}{2}} +\ket{\frac{1}{2}\frac{1}{2},-\frac{1}{2}}\right) . \ \\
   \end{aligned}
\end{equation}
\noindent
The kets of the above equations represent $\ket{I^{\eta K^*} I^{\eta K^*}_z,I_z^{\bar{K}^*}}$ ($\ket{I^{\eta \bar{K}^*} I^{\eta \bar{K}^*}_z,I_z^{K^*}}$). So, we have 
\begin{equation}
   \begin{aligned}
      t_1 &=\bra{A_1}\hat{t}_1\ket{A_1} = \frac{1}{2}t^{I=\frac{1}{2}}_{\eta K^* \to \eta K^*}+\frac{1}{2}t^{I=\frac{1}{2}}_{\eta K^* \to \eta K^*} \\
      &=t^{I=\frac{1}{2}}_{\eta K^* \to \eta K^*}, \\
      t_2 &=\bra{A_2}\hat{t}_1\ket{A_2} = \frac{1}{2}t^{I=\frac{1}{2}}_{\eta \bar{K}^* \to \eta \bar{K}^*}+\frac{1}{2}t^{I=\frac{1}{2}}_{\eta \bar{K}^* \to \eta \bar{K}^*}\\
      &=t^{I=\frac{1}{2}}_{\eta \bar{K}^* \to \eta \bar{K}^*}.
  \end{aligned}
\end{equation}

\begin{table}[htbp]
	\centering
	\renewcommand\arraystretch{1.5}
	\caption{Three body single scattering amplitudes in terms of the unitarized two-body scattering amplitudes. Here, $I$ denotes the total isospin of the discussed three-body sysmtems.}
	\label{Tab:C}
	\begin{tabular}{c|c|p{7mm}<{\centering}|c}
		\toprule[1pt]
		\toprule[1pt]
		\thead{Fixed-center \\ $K^*\bar{K}^*$}  & \thead{Three-body\\ systems} & \thead{$I$} & \\
		\hline
		\multirow{6}{*}{\thead{\\\\\\\\\\\\\\\\$a_0(1780)$}}&$\eta K^*\bar{K}^*$&1 &\thead{$t_1=t_{\eta K^* \to \eta K^*}$\\[0.5em]$t_2=t_{\eta \bar{K}^* \to \eta \bar{K}^*}$}\\[0.5em]
		\cline{2-4}
		&\multirow{3}{*}{\thead{\\\\\\$\pi K^*\bar{K}^*$}}&0&\thead{$t_1=t_{\pi K^* \to \pi K^*}^{I=\frac{1}{2}}$\\[0.5em]$t_2=t_{\pi \bar{K}^* \to \pi \bar{K}^*}^{I=\frac{1}{2}}$}\\[0.5em]
		\cline{3-4}
		&&1&\thead{$t_1=\frac{2}{3}t_{\pi K^* \to \pi K^*}^{I=\frac{1}{2}}+\frac{1}{3}t_{\pi K^* \to \pi K^*}^{I=\frac{3}{2}}$\\[0.5em]$t_2=\frac{2}{3}t_{\pi \bar{K}^* \to \pi \bar{K}^*}^{I=\frac{1}{2}}+\frac{1}{3}t_{\pi \bar{K}^* \to \pi \bar{K}^*}^{I=\frac{3}{2}}$}\\[0.5em]
		\cline{3-4}
		&&2&\thead{$t_1=t_{\pi K^* \to \pi K^*}^{I=\frac{3}{2}}$\\[0.5em]$t_2=t_{\pi \bar{K}^* \to \pi \bar{K}^*}^{I=\frac{3}{2}}$}\\[0.5em]
		\cline{2-4}
		&\multirow{2}{*}{\thead{\\$K K^*\bar{K}^*$}}&$\frac{1}{2}$&\thead{$t_1=\frac{3}{4}t_{KK^* \to KK^*}^{I=0}+\frac{1}{4}t_{KK^* \to KK^*}^{I=1}$\\[0.5em]$t_2=\frac{3}{4}t_{K\bar{K}^* \to K\bar{K}^*}^{I=0}+\frac{1}{4}t_{K\bar{K}^* \to K\bar{K}^*}^{I=1}$}\\[0.5em]
		\cline{3-4}
		&&$\frac{3}{2}$&\thead{$t_1=t_{KK^* \to KK^*}^{I=1}$\\[0.5em]$t_2=t_{K\bar{K}^* \to K\bar{K}^*}^{I=1}$}\\[0.5em]
		\cline{1-4}
		\multirow{3}{*}{\thead{\\\\\\$f_2'(1525)$}}&$\eta K^*\bar{K}^*$&0&\thead{$t_1=t_{\eta K^* \to \eta K^*}$\\[0.5em]$t_2=t_{\eta \bar{K}^* \to \eta \bar{K}^*}$}\\[0.5em]
		\cline{2-4}
		&$\pi K^*\bar{K}^*$&1&\thead{$t_1=\frac{1}{3}t_{\pi K^* \to \pi K^*}^{I=\frac{1}{2}}+\frac{2}{3}t_{\pi K^* \to \pi K^*}^{I=\frac{3}{2}}$\\[0.5em]$t_2=\frac{1}{3}t_{\pi \bar{K}^* \to \pi \bar{K}^*}^{I=\frac{1}{2}}+\frac{2}{3}t_{\pi \bar{K}^* \to \pi \bar{K}^*}^{I=\frac{3}{2}}$}\\[0.5em]
		\cline{2-4}
		&$KK^*\bar{K}^*$&$\frac{1}{2}$&\thead{$t_1=\frac{1}{4}t_{KK^* \to KK^*}^{I=0}+\frac{3}{4}t_{KK^* \to KK^*}^{I=1}$\\[0.5em]$t_2=\frac{1}{4}t_{K\bar{K}^* \to K\bar{K}^*}^{I=0}+\frac{3}{4}t_{K\bar{K}^* \to K\bar{K}^*}^{I=1}$}\\[0.5em]
		\bottomrule[1pt]
		\bottomrule[1pt]
	\end{tabular}
\end{table}

Using the same procedures, one can easily obtain all the amplitudes for the single-scattering contribution in the present calculation which are shown in table~\ref{Tab:C} for the case of $\eta/\pi/ K$-$(K^*\bar{K}^*)_{a_0(1780)}$ and $\eta/\pi/ K$-$(K^*\bar{K}^*)_{f'_2(1525)}$ configurations with different total isospins.

On the other hand, following the approach developed in Refs.~\cite{Yamagata-Sekihara:2010muv,Roca:2010tf}, we need to give a weight to the two-body scattering amplitudes $t_1$ and $t_2$ so that we have the correct normalization for the meson fields. This is achieved by replacing
\begin{equation}
   \begin{aligned}
      t_1 \to \tilde{t}_1=\frac{M_{B^*}}{m_{K^*}}t_1,~~~
      t_2 \to \tilde{t}_2=\frac{M_{B^*}}{m_{\bar{K}^*}}t_2.
   \end{aligned}
\end{equation}

In addition, we also consider the effect of single-scattering above the mass threshold of particle 3 and the bound state $B^*$. Following Refs.~\cite{Roca:2011br,Zhang:2016bmy}, we need to project the form factor into the $s$-wave. Then we have~\footnote{The form factor $FFS$ was taken to be unity in Ref.~\cite{Shen:2022etd}, since for the $\eta (K^*\bar{K}^*)_{f_0(1710)}$ and $K (K^*\bar{K}^*)_{f_0(1710)}$ systems, only states below threshold were found. While for the $\pi (K^*\bar{K}^*)_{f_0(1710)}$ system, there are uncertainties of about 20 MeV for the peak position of the modulus squared of the three-body scattering amplitudes, which is a small effect. Therefore, the main conclusions there are unchanged when the form factor $FFS$ is taken into account.}
\begin{equation}
FFS_1(s) = FFS_2(s) =\frac{1}{2}\int^{+1}_{-1} F_{B^*} (k_1)d({\rm cos} \theta) ,   \label{eq:FFS}
\end{equation}
\noindent
with $	k_1 = \frac{k}{2} \sqrt{2(1 - {\rm cos} \theta)}$, and 
\begin{eqnarray}
	k &=&
	\frac{\sqrt{(s-(M_{B^*}+m_3)^2)(s-(M_{B^*}-m_3)^2)}}{2\sqrt{s}} ,
	\label{eq:k} \nonumber
\end{eqnarray}
for $\sqrt{s} \ge M_{B^*}+m_3 $. Otherwise, $k=0$. Then, the Eq.~\eqref{eq:FCA} can be rewritten as
\begin{align}
   \tilde{T}_1 &=FFS_1\cdot \tilde{t}_1+\tilde{t}_1G_0 \tilde{t}_2 = (FFS_1-1)\tilde{t}_1+T_1,\\
   \tilde{T}_2 &=(FFS_2-1)\tilde{t}_2+T_2 ,\\
   T&= \tilde{T}_1 + \tilde{T}_2  =  \frac{\tilde{t}_1+\tilde{t}_2+2\tilde{t}_1\tilde{t}_2G_0}{1-\tilde{t}_1\tilde{t}_2G_0^2}+(FFS_1-1)\tilde{t}_1 \notag \\ &\quad+(FFS_2-1)\tilde{t}_2 .
   \label{eq:T-FFS}
\end{align}
The analysis of the $\pi/\eta/K-(K^*\bar{K}^*)_{a_0(1780)/f'_2(1525)}$ scattering amplitudes $T$ will allow us to study dynamically generated resonances. 

It is worth noting that the total three-body scattering amplitude $T$ is a function of the total invariant mass $\sqrt{s}$ of the three-body system. While the two-body scattering amplitudes $t_1$ and $t_2$ depend on the invariant masses $\sqrt{s_1}$ and $\sqrt{s_2}$, which are the invariant masses of $\eta$ ($\pi$ or $K$) and the particle $K^*$ ($\bar{K}^*$) within the bound state of $a_0(1780)$ or $f_2'(1525)$. The $s_1$ and $s_2$ are: $s_1 = s_2 = m_3^2+m_{K^*}^2 + (s-m_3^2-M_{B^*}^2)/2 $.

\section{Numerical results}~\label{numericalresults}

In this section, we will show the theoretical numerical results obtained for the scattering amplitude modulus square of the three-body systems $\eta K^*\bar{K}^*$, $\pi K^*\bar{K}^*$, and $KK^*\bar{K}^*$, respectively, and we evaluate the three-body scattering amplitude $T$ and associate the peaks or bumps in the modulus squared of $|T|^2$ with resonances.

\subsection{There-body system with the $K^*\bar{K}^*$ subsystem as $a_0(1780)$}

\subsubsection{Three-body $\eta (K^*\bar{K}^*)_{a_0(1780)}$ system}~\label{sec:etaKstarKbarstar}

For the $\eta K^*\bar{K}^*$ system, its total isospin is one because $K^*\bar{K}^*$ has isospin one and $\eta$ meson is zero. In order to obtain the three-body scattering amplitudes, one needs to obtain these two-body scattering amplitudes $t_1$ and $t_2$. While $t_1$ and $t_2$ can be obtained with these scattering amplitudes of $\eta K^*$ and $\eta \bar{K}^*$, which are taken from these previous works as in Refs.~\cite{Lutz:2003fm,Roca:2005nm,Geng:2006yb}. Furthermore, we also consider the width of the vector meson~\cite{Geng:2006yb} and the effect of the $\eta'$ meson as done in Refs.~\cite{Guo:2005wp,Sun:2022cxf}. With these model parameters as used in Refs.~\cite{Roca:2005nm,Geng:2006yb}, the two-body scattering amplitude of $t_{\eta K^* \to \eta K^*}$ can be easily obtained and one can find that the interaction between $\eta$ and $K^*$ is strong.

In fact, it is expected that the $\eta K^*\bar{K}^*$ three-body system could be bound, since these interactions between $\eta$, $K^*$ and $\bar{K}^*$ are all strong and attractive. The modulus squared scattering amplitude $T_{\eta (K^*\bar{K}^*)_{a_0(1780)} \to \eta (K^*\bar{K}^*)_{a_0(1780)}}$ is shown in Fig.~\ref{fig:T} (a), showing a clear peak structure around 2122 MeV, which can be associated with the $\pi (2070)$. There is some evidence for this state in a combined partial wave analysis of $p \bar{p}$ annihilation channels~\cite{Anisovich:2001pn}. It is also cited as further state in RPP~\cite{ParticleDataGroup:2022pth}, and its mass and width are about $2070 \pm 35$ MeV and $310^{+100}_{-50}$ MeV, respectively. Here, we explain the $\pi(2070)$ meson as a $\eta (K^*\bar{K}^*)_{a_0(1780)}$ molecular state. Improved experimental data are desirable to draw more firm conclusions. 


\subsubsection{Three-body $\pi (K^*\bar{K}^*)_{a_0(1780)}$ system}

In the case of a three-body $\pi K^*\bar{K}^*$ system, its total isospin could be zero, one or two. We need two-body coupled-channels scattering amplitudes $t_{\pi K^* \to \pi K^*}$ in the $I_{\pi K^*} = \frac{1}{2}$ and $\frac{3}{2}$ sectors. Within the formula and theoretical parameters as in Refs.~\cite{Geng:2006yb}, one can easily obtain the two-body scattering amplitude $t_{\pi K^* \to \pi K^*}$ in coupled channels. And then we can calculate the three-body scattering amplitude $T_{\pi(K^*\bar{K}^*)_{a_0(1780)} \to \pi(K^*\bar{K}^*)_{a_0(1780)}}$.

\begin{figure*}[htbp]
    \centering
    \includegraphics[scale=0.6]{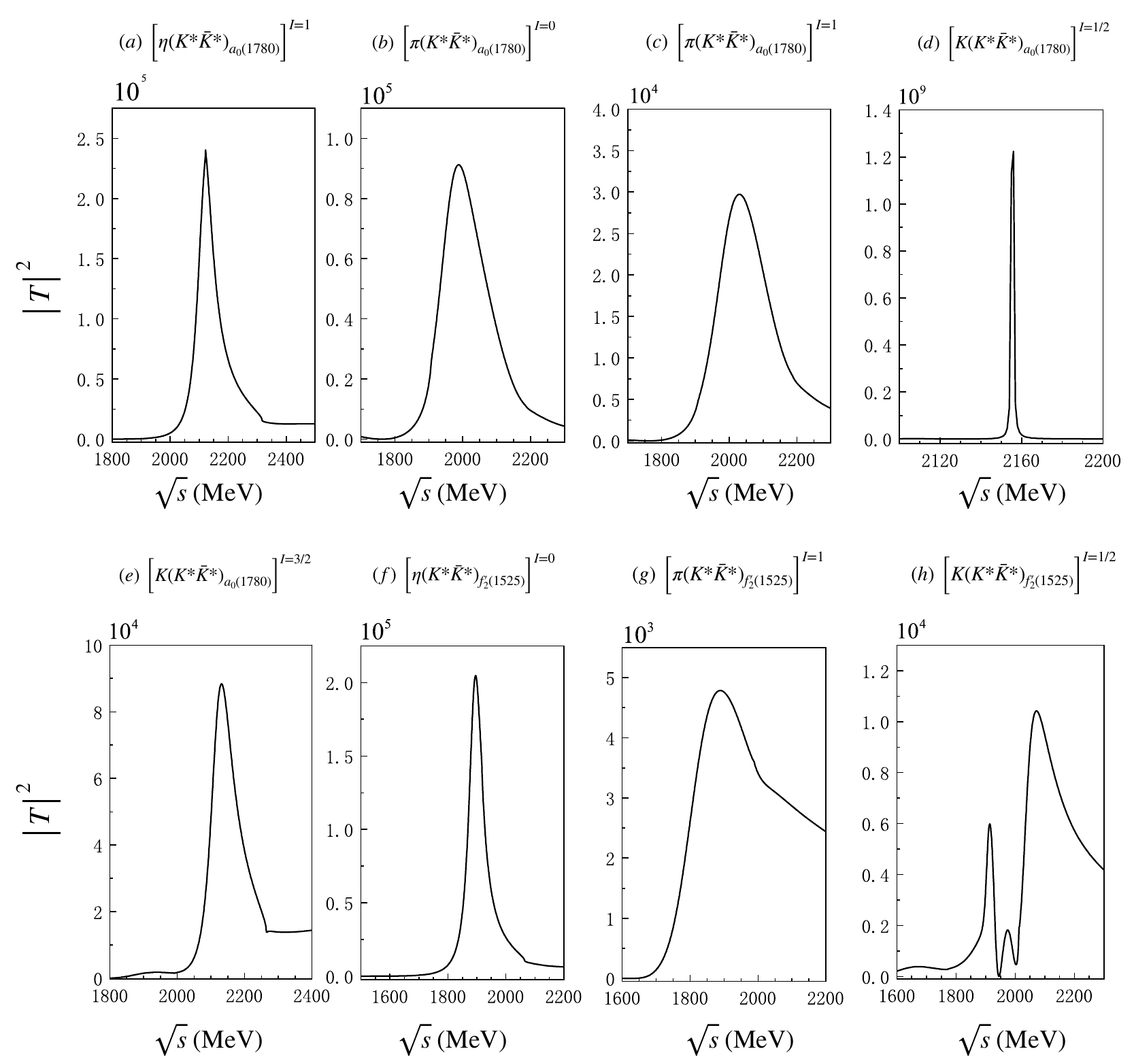}
    \caption{Modulus squard scattering amplitude $|T|^2$ for three-body system $\eta / \pi/ K(K^*\bar{K}^*)_{a_0(1780)\left(f_2'(1525)\right)}$ with $\Lambda=1100(1009)$ MeV. }
    \label{fig:T}
\end{figure*}

For the case of $I=0$, one can see a bump structure located at 1988 MeV in Fig.~\ref{fig:T} (b), which can be interpreted as a $\eta$ excited state $\eta(2010)$ in RPP~\cite{ParticleDataGroup:2022pth}. It is found in a combining fit to data on $p\bar{p}$ annihilation~\cite{Anisovich:2000ut}, with mass $2010^{+35}_{-60}$ MeV and width $270\pm60$ MeV. 

For the case of $I=1$, as shown in Fig.~\ref{fig:T} (c), a bump structures is located around 2030 MeV, and it may be associated with the $\pi(2070)$ state, which was not quoted in the summary table of RPP~\cite{ParticleDataGroup:2022pth}. However, the $\pi(2070)$ state was required in a combined partial wave analysis of the $\bar{p}p$ annihilation channels as studied in Ref.~\cite{Anisovich:2000ut}. It is hoped that further experimental measurements will test this prediction.

Finally, no special structures are found for the case of $I=2$, because the interactions of $\pi K^*$ and $\pi \bar{K}^*$ in the isospin $I=\frac{3}{2}$ sector are rather small.

\subsubsection{Three-body $K (K^*\bar{K}^*)_{a_0(1780)}$ system}

For the three-body $K(K^*\bar{K}^*)_{a_0(1780)}$ system, its total isospin can be either $I=\frac{1}{2}$ or $I=\frac{3}{2}$. The modulus squared of the scattering amplitudes $|T_{K (K^*\bar{K}^*)_{a_0(1780)} \to K (K^*\bar{K}^*)_{a_0(1780)}}|^2$ are shown in Fig.~\ref{fig:T}. For the case of $I=\frac{1}{2}$, a resonance can be found around 2156 MeV as shown in Fig.~\ref{fig:T} (d). It should be mixed with a structure that is found in Ref.~\cite{Shen:2022etd} with mass around $2130$ MeV. 

For the case of $I=\frac{3}{2}$, one can also find a peak structure around 2130 MeV in Fig.~\ref{fig:T} (e). We look forward to observing this low-lying exotic mesonic state with isospin $I=\frac{3}{2}$ in future experimental measurements.

\subsubsection{Effects from the mass of $a_0(1780)$}

Since the scalar meson $a_0(1780)$ is not well established in the RPP~\cite{ParticleDataGroup:2022pth}, and its mass has some uncertainties, we should also consider the effects of the mass of $a_0(1780)$. We do this by adjusting the cutoff $\Lambda$ from 1000 to 1500 MeV. For different cutoffs $\Lambda$, we can get different masses of $a_0(1780)$ and peak positions of the squared of the three-body scattering amplitude, which are shown in Table~\ref{tab:summary}. We find that with these numerical results for the mass of $a_0(1780)$ one can always find peak structures in the squared of the three-body scattering amplitude. For the $\eta K^*\bar{K}^*$ and $\pi K^*\bar{K}^*$ systems with $I=1$, they are associated with $\pi(2070)$, and for the $\pi K^*\bar{K}^*$ sysmtem with the total isospin $I=0$, they can be associated with $\eta(2010)$. We also find two unobserved states in the $K (K^*\bar{K}^*)_{a_0(1780)}$ system, one of which carries exotic quantum numbers $I(J^P) = \frac{3}{2}(0^-)$. 

\begin{table}[htbp]
   \centering
   \renewcommand\arraystretch{2}
   \tabcolsep=0.2cm
   \caption{The calculated mass of $a_0(1780)$ and the peak positions ($M_{\eta K^*\bar{K}^*}$, $M_{\pi K^*\bar{K}^*}$, and $M_{K K^*\bar{K}^*}$) of squared of three-body scattering amplitude corresponding to different values of the cutoff parameter $\Lambda$. Here, all values are in units of MeV.}
   \begin{tabular}{c|c|c|c|c|c|c}
      \toprule[1pt]
      \toprule[1pt]
      $\Lambda$&1000&1100&1200&1300&1400&1500 \\
      \hline
      $M_{a_0(1780)}$&1779&1769&1759&1749&1740&1731 \\
      \hline
      $M_{[\eta K^*\bar{K}^*]^{I=1}}$ &2135&2122&2120&2098&2086&2075 \\
      \hline
      $M_{[\pi K^*\bar{K}^*]^{I=0}}$&\thead{2015}&\thead{1988}&\thead{1968}&\thead{1952}&\thead{1937}&\thead{1924}\\
      \hline
      $M_{[\pi K^*\bar{K}^*]^{I=1}}$&\thead{2048}&\thead{2030}&\thead{2013}&\thead{1997}&\thead{1985}&\thead{1972} \\
      \hline
    $M_{[KK^*\bar{K}^*]^{I=\frac{1}{2}}}$&2161&2156&2150&2145&2141&2138\\
      \hline
      $M_{[KK^*\bar{K}^*]^{I=\frac{3}{2}}}$&2131&2132&2134&2138&2145&2152\\
      \bottomrule[1pt]
      \bottomrule[1pt]
   \end{tabular}
   \label{tab:summary}
\end{table}

\subsection{There-body system with $K^*\bar{K}^*$ as $f_2'(1525)$}

\subsubsection{Three-body $\eta (K^*\bar{K}^*)_{f'_2(1525)}$ system}

We show the modulus squared of the scattering amplitude for $T_{\eta (K^*\bar{K}^*)_{f'_2(1525)} \to \eta (K^*\bar{K}^*)_{f'_2(1525)}}$ in Fig.~\ref{fig:T} (f) with $\Lambda=1009$ MeV. A distinct peak structure is found around 1896 MeV. It can be interpreted as $\eta_2(1870)$ with quantum numbers $J^{PC} = 2^{-+}$. Although the $\eta_2(1870)$ meson has been confirmed by several experiments~\cite{ParticleDataGroup:2022pth}, there are difference theoretical explanations for this controversial state. It is very interesting that the $\eta_2(1870)$ is labeled as a no-$q\bar{q}$ state in RPP~\cite{ParticleDataGroup:2022pth}. Its experimentally measured mass is between 1835 MeV~\cite{WA102:1999ybu} and 1881 MeV~\cite{CrystalBall:1991zkb}, and the average mass is $1842\pm8$ MeV and its average width is $225\pm 14$ MeV~\cite{ParticleDataGroup:2022pth}. In addition, the $\eta_2(1870)$ state has been interpreted as a hybrid state, mixed by the $s\bar{s}(^1D_2)$ and $q\bar{q}(2^1D_2)$ states. Its mass, calculated by different models, is listed in Table~\ref{tab:mass-eta1870}, where GI stands for the Godfrey-Isgur quark model and VFV for the Vijande-Fernandez-Valcarce quark model.


\begin{table}[htbp]
   \centering
   \renewcommand\arraystretch{1.5}
   \tabcolsep=0.13cm
   \caption{Mass of the $\eta_2(1870)$ meson calculated by different model. Here, all masses are given in units of MeV.}
   \begin{tabular}{c|c|c|c|c|c|c|c}
      \toprule[1pt]
      \toprule[1pt]
      \multirow{2}{*}{}&\multirow{2}{*}{\thead{This \\work}}&\multirow{2}{*}{\thead{Hybrid \\ state\\~\cite{Isgur:1984bm}}}&\multicolumn{2}{c|}{$s\bar{s}(^1D_2) $}&\multicolumn{2}{c|}{$q\bar{q}(2^1D_2)$}&\multirow{2}{*}{\thead{RPP\\~\cite{ParticleDataGroup:2022pth}}}\\
      \cline{4-7}
      &&&\thead{GI\\~\cite{Godfrey:1985xj}}&\thead{VFV\\~\cite{Vijande:2004he}}&\thead{GI\\~\cite{Godfrey:1985xj}}&\thead{VFV\\~\cite{Vijande:2004he}}& \\
      \hline
      Mass&1896&1900&1890&1853&2130&1863&$1842\pm8$\\
      \bottomrule[1pt]
      \bottomrule[1pt]
   \end{tabular}
      \label{tab:mass-eta1870}
\end{table}

A major problem with the $\eta_2(1870)$ meson is that it did not appear in the $K^*\bar{K}$ decay channel, and it is difficult to confirm the $\eta_2(1870)$ state as a conventional $s\bar{s}$ state~\cite{Wang:2014sea} (see this reference for more details). On the other hand, if the $\eta_2(1870)$ state is explained as a $q\bar{q}(2 ^1 D_2) $ state, its mass is too low and the theoretical branching ratio is~\cite{Li:2009rka}
\begin{align}
   \frac{ \Gamma(\eta_2(1780)\to K^*\bar{K} )}{\Gamma(\eta_2(1780)\to f_2(1270)\eta)} \approx 1 .
\end{align}
However, the $\eta_2(1870)$ was not seen in the $\bar{K}K^*$ channel, and the $\eta_2(1870) \to f_2(1270)\eta$ decay mode has been observed in many experiments. In this work, it is found that the $\eta_2(1870)$ can be interpreted as a $\eta$-$(K^*\bar{K}^*)_{f_2'(1525)}$ three-body resonance.

\subsubsection{Three-body $\pi (K^*\bar{K}^*)_{f'_2(1525)}$ system}

The modulus squared of the scattering amplitude for $T_{\pi (K^*\bar{K}^*)_{f_2'(1525)} \to \pi (K^*\bar{K}^*)_{f_2'(1525)}}$ is shown in Fig.~\ref{fig:T} (g) with $\Lambda=1009$ MeV. A bump structure appears around $1900$ MeV and it can be associated with the $\pi_2(1880)$ meson. The $\pi_2(1880)$ is well established in RPP~\cite{ParticleDataGroup:2022pth}. Its average mass is $1876^{+26}_{-5}$ MeV and its width is $237^{+33}_{-30}$ MeV. However, its mass is too light to be the first radial excitation of $\pi_2(1670)$~\cite{Wang:2014sea}.  In this work, we find that  $\pi_2(1880)$ can be interpreted as a $\pi$-$(K^*\bar{K}^*)_{f_2'(1525)}$ three-body resonance. On the other hand, the $\pi_2(1880)$ meson was interpreted as hybrid~\cite{Anisovich:2001hj,Barnes:1995hc} and $q\bar{q}(2^1D_2)$~\cite{Godfrey:1985xj}. The mass obtained for $\pi_2(1880)$ from different models are listed Table~\ref{tab:mass-pi1880}.


\begin{table}[htbp]
   \centering
   \renewcommand\arraystretch{1.5}
   \tabcolsep=0.13cm
   \caption{Mass of $\pi_2(1880)$ calculated by different model. Here, all mass values listed in the second row are in units of MeV.}
   \begin{tabular}{c|c|c|c|c}
      \toprule[1pt]
      \toprule[1pt]
      Model&This work&Hybrid state~\cite{Barnes:1995hc}&$q\bar{q}(2^1D_2)$~\cite{Godfrey:1985xj}&RPP~\cite{ParticleDataGroup:2022pth}\\
     \hline
       Mass&1889&1800-1900&2130&$1876^{+26}_{-5}$\\
      \bottomrule[1pt]
      \bottomrule[1pt]
   \end{tabular}
   \label{tab:mass-pi1880}
\end{table}

\subsubsection{Three-body $K (K^*\bar{K}^*)_{f'_2(1525)}$ system}

For the $K (K^*\bar{K}^*)_{f'_2(1525)}$ system, the quantum number is $J^{P} = 2^{-}$ and the isospin is $I=\frac{1}{2}$. We will thereefore study the excited $K_2$ states. There is very little information available on the $K_2$ states. As shown in RPP~\cite{ParticleDataGroup:2022pth}, only four $K_2$ states with spin-parity $J^P =2^-$ are cataloged, which are $K_2(1580)$, $K_2(1770)$, $K_2(1820)$, and $K_2(2250)$. Most of their properties are unknown~\cite{ParticleDataGroup:2022pth}.

In Fig.~\ref{fig:T} (h), we show the modulus squared scattering amplitude for $T_{K(K^*\bar{K}^*)_{f_2'(1525)} \to K(K^*\bar{K}^*)_{f_2'(1525)}}$. The numerical results are calculated with $\Lambda=1009$ MeV. It can be seen that there are three distinct peaks located at 1914, 1975, and 2072 MeV. However, none of them can be associated with the four excited $K_2$ states quoted in RPP~\cite{ParticleDataGroup:2022pth} as discussed above. Further experimental and theoretical works is needed in this direction.


\section{Summary}~\label{Summary}

Within the framework of the fixed-center approximation, we further study the $\eta K^*\bar{K}^*$, $\pi K^*\bar{K}^*$ and $KK^*\bar{K}^*$ three-body system where we view the $K^*\bar{K}^*$ subsystem as scalar meson $a_0(1780)$ and tensor meson $f_2'(1525)$. In terms of the two-body interactions, $\eta (\pi / K) K^*(\bar{K}^*)$ and $K^*\bar{K}^*$ provided by the chiral unitary approach, we describe the $\eta/ \pi / K$-$(K^*\bar{K}^*)$ systems by using the FCA. By analysis of the $\eta/ \pi / K$-$(K^*\bar{K}^*)_{a_0(1780)/f'_2(1525)}$ scattering amplitudes, one can study those dynamically generated resonances from the above three-body systems. It is found that the $\eta(2010)$ meson can be interpreted as $\pi (K^*\bar{K}^*)_{a_0(1780)}$ with $I=0$, and the $\pi(2070)$ can be explained with the $\pi (K^*\bar{K}^*)_{a_0(1780)}$ with $I=1$ and $\eta (K^*\bar{K}^*)_{a_0(1780)}$. Two resonances with masses around 2150 MeV are predicted in $K(K^*\bar{K}^*)_{a_0(1780)}$ with $I=\frac{1}{2}$ and $\frac{3}{2}$. It is important to observe such an exotic light flavor state with isospin $I=\frac{3}{2}$ by future experiments.

\begin{table}[htbp]
   \centering
   \renewcommand\arraystretch{1.7}
   \tabcolsep=0.15cm
   \caption{Summary about the theoretical results obtained in this work for the $\eta K^*\bar{K}^*$, $\pi K^*\bar{K}^*$ and $KK^*\bar{K}^*$ three-body systems. The question mark "?" stands for a non-cataloged state in the Review of particle physics~\cite{ParticleDataGroup:2022pth}.}
   \begin{tabular}{c|c|c|c|c}
      \toprule[1pt]
      \toprule[1pt]
      Systems& $I^G(J^{PC})$&States&\thead{Mass (MeV) \\ this work}&\thead{Mass (MeV) \\ RPP~\cite{ParticleDataGroup:2022pth}}\\
      \hline
      \multirow{2}{*}{$\eta K^*\bar{K}^*$}&$1^-(0^{-+})$&$\pi(2070)$&$2122$&$2070\pm35$\\
      \cline{2-5}
      &$0^+(2^{-+})$&$\eta_2(1870)$&$1896$&$1842\pm8$\\
      \hline
      \multirow{3}{*}{$\pi K^*\bar{K}^*$}&$0^+(0^{-+})$&$\eta(2010)$&$1988$&$2010^{+35}_{-60}$\\
      \cline{2-5}
      &$1^-(0^{-+})$&$\pi(2070)$&$2030$&$2070\pm35$\\
      \cline{2-5}
      &$1^-(2^{-+})$&$\pi_2(1880)$&$1889$&$1876^{+26}_{-5}$\\
      \hline
      \multirow{5}{*}{$KK^*\bar{K}^*$}&$\frac{1}{2}(0^-)$&?&$2156$&$?$\\
      \cline{2-5}
      &$\frac{3}{2}(0^-)$&?&$2132$&?\\
      \cline{2-5}
      &\multirow{3}{*}{$\frac{1}{2}(2^-)$}&?&$1914$&?\\
      \cline{3-5}
      & &?&$1975$&?\\
      \cline{3-5}
      & &?&$2072$&?\\
      \bottomrule[1pt]
      \bottomrule[1pt]
   \end{tabular}
   \label{tab:summary-all}
\end{table}

Furthermore,  the generated resonances with quantum numbers $J^{P} = 2^{-}$ can be unambiguously assigned to experimental states. The $\eta_2(1870)$ meson can be interpreted as $\eta( K^*\bar{K}^*)_{f'_2(1525)}$, while $\pi_2(1880)$ can be interpreted as $\pi (K^*\bar{K}^*)_{f'_2(1525)}$. This assignment provides a natural explanation to these states. Actually the possibility of providing a theoretical explanation of such resonances was the main motivation for our
study since its description is clearly out of the scope of the classical $q\bar{q}$ model.

Finally, we summarize the theoretical results obtained here about $\eta K^*\bar{K}^*$, $\pi K^*\bar{K}^*$, and $KK^*\bar{K}^*$ three-body system in Table~\ref{tab:summary-all}. It is expected these theoretical calculations could be tested by future experiments, such as the BESIII, BellII, and LHCb.

\section*{Acknowledgements}

We warmly thank Prof. Li-Sheng Geng for useful comments and discussions. This work is partly supported by the National Natural Science Foundation of China under Grant Nos. 12075288, 12247101 and 12335001. It is also supported by the Youth Innovation Promotion Association CAS. LX is also supported by the China National Funds for Distinguished Young Scientists under Grant No. 11825503, National Key Research and Development Program of China under Contract No. 2020YFA0406400, the 111 Project under Grant No. B20063, the Fundamental Research Funds for the Central Universities, the project for top-notch innovative talents of Gansu province. 

\vfil

\section*{}
\normalem
\bibliographystyle{apsrev4-1.bst}
\bibliography{reference.bib}

\begin{thebibliography}{98}%
\makeatletter
\providecommand \@ifxundefined [1]{%
 \@ifx{#1\undefined}
}%
\providecommand \@ifnum [1]{%
 \ifnum #1\expandafter \@firstoftwo
 \else \expandafter \@secondoftwo
 \fi
}%
\providecommand \@ifx [1]{%
 \ifx #1\expandafter \@firstoftwo
 \else \expandafter \@secondoftwo
 \fi
}%
\providecommand \natexlab [1]{#1}%
\providecommand \enquote  [1]{``#1''}%
\providecommand \bibnamefont  [1]{#1}%
\providecommand \bibfnamefont [1]{#1}%
\providecommand \citenamefont [1]{#1}%
\providecommand \href@noop [0]{\@secondoftwo}%
\providecommand \href [0]{\begingroup \@sanitize@url \@href}%
\providecommand \@href[1]{\@@startlink{#1}\@@href}%
\providecommand \@@href[1]{\endgroup#1\@@endlink}%
\providecommand \@sanitize@url [0]{\catcode `\\12\catcode `\$12\catcode
  `\&12\catcode `\#12\catcode `\^12\catcode `\_12\catcode `\%12\relax}%
\providecommand \@@startlink[1]{}%
\providecommand \@@endlink[0]{}%
\providecommand \url  [0]{\begingroup\@sanitize@url \@url }%
\providecommand \@url [1]{\endgroup\@href {#1}{\urlprefix }}%
\providecommand \urlprefix  [0]{URL }%
\providecommand \Eprint [0]{\href }%
\providecommand \doibase [0]{http://dx.doi.org/}%
\providecommand \selectlanguage [0]{\@gobble}%
\providecommand \bibinfo  [0]{\@secondoftwo}%
\providecommand \bibfield  [0]{\@secondoftwo}%
\providecommand \translation [1]{[#1]}%
\providecommand \BibitemOpen [0]{}%
\providecommand \bibitemStop [0]{}%
\providecommand \bibitemNoStop [0]{.\EOS\space}%
\providecommand \EOS [0]{\spacefactor3000\relax}%
\providecommand \BibitemShut  [1]{\csname bibitem#1\endcsname}%
\let\auto@bib@innerbib\@empty
\bibitem [{\citenamefont {Liu}(2014)}]{Liu:2013waa}%
  \BibitemOpen
  \bibfield  {author} {\bibinfo {author} {\bibfnamefont {X.}~\bibnamefont
  {Liu}},\ }\href {\doibase 10.1007/s11434-014-0407-2} {\bibfield  {journal}
  {\bibinfo  {journal} {Chin. Sci. Bull.}\ }\textbf {\bibinfo {volume} {59}},\
  \bibinfo {pages} {3815} (\bibinfo {year} {2014})},\ \Eprint
  {http://arxiv.org/abs/1312.7408} {arXiv:1312.7408 [hep-ph]} \BibitemShut
  {NoStop}%
\bibitem [{\citenamefont {Hosaka}\ \emph {et~al.}(2016)\citenamefont {Hosaka},
  \citenamefont {Iijima}, \citenamefont {Miyabayashi}, \citenamefont {Sakai},\
  and\ \citenamefont {Yasui}}]{Hosaka:2016pey}%
  \BibitemOpen
  \bibfield  {author} {\bibinfo {author} {\bibfnamefont {A.}~\bibnamefont
  {Hosaka}}, \bibinfo {author} {\bibfnamefont {T.}~\bibnamefont {Iijima}},
  \bibinfo {author} {\bibfnamefont {K.}~\bibnamefont {Miyabayashi}}, \bibinfo
  {author} {\bibfnamefont {Y.}~\bibnamefont {Sakai}}, \ and\ \bibinfo {author}
  {\bibfnamefont {S.}~\bibnamefont {Yasui}},\ }\href {\doibase
  10.1093/ptep/ptw045} {\bibfield  {journal} {\bibinfo  {journal} {PTEP}\
  }\textbf {\bibinfo {volume} {2016}},\ \bibinfo {pages} {062C01} (\bibinfo
  {year} {2016})},\ \Eprint {http://arxiv.org/abs/1603.09229} {arXiv:1603.09229
  [hep-ph]} \BibitemShut {NoStop}%
\bibitem [{\citenamefont {Chen}\ \emph {et~al.}(2016)\citenamefont {Chen},
  \citenamefont {Chen}, \citenamefont {Liu},\ and\ \citenamefont
  {Zhu}}]{Chen:2016qju}%
  \BibitemOpen
  \bibfield  {author} {\bibinfo {author} {\bibfnamefont {H.-X.}\ \bibnamefont
  {Chen}}, \bibinfo {author} {\bibfnamefont {W.}~\bibnamefont {Chen}}, \bibinfo
  {author} {\bibfnamefont {X.}~\bibnamefont {Liu}}, \ and\ \bibinfo {author}
  {\bibfnamefont {S.-L.}\ \bibnamefont {Zhu}},\ }\href {\doibase
  10.1016/j.physrep.2016.05.004} {\bibfield  {journal} {\bibinfo  {journal}
  {Phys. Rept.}\ }\textbf {\bibinfo {volume} {639}},\ \bibinfo {pages} {1}
  (\bibinfo {year} {2016})},\ \Eprint {http://arxiv.org/abs/1601.02092}
  {arXiv:1601.02092 [hep-ph]} \BibitemShut {NoStop}%
\bibitem [{\citenamefont {Richard}(2016)}]{Richard:2016eis}%
  \BibitemOpen
  \bibfield  {author} {\bibinfo {author} {\bibfnamefont {J.-M.}\ \bibnamefont
  {Richard}},\ }\href {\doibase 10.1007/s00601-016-1159-0} {\bibfield
  {journal} {\bibinfo  {journal} {Few Body Syst.}\ }\textbf {\bibinfo {volume}
  {57}},\ \bibinfo {pages} {1185} (\bibinfo {year} {2016})},\ \Eprint
  {http://arxiv.org/abs/1606.08593} {arXiv:1606.08593 [hep-ph]} \BibitemShut
  {NoStop}%
\bibitem [{\citenamefont {Lebed}\ \emph {et~al.}(2017)\citenamefont {Lebed},
  \citenamefont {Mitchell},\ and\ \citenamefont {Swanson}}]{Lebed:2016hpi}%
  \BibitemOpen
  \bibfield  {author} {\bibinfo {author} {\bibfnamefont {R.~F.}\ \bibnamefont
  {Lebed}}, \bibinfo {author} {\bibfnamefont {R.~E.}\ \bibnamefont {Mitchell}},
  \ and\ \bibinfo {author} {\bibfnamefont {E.~S.}\ \bibnamefont {Swanson}},\
  }\href {\doibase 10.1016/j.ppnp.2016.11.003} {\bibfield  {journal} {\bibinfo
  {journal} {Prog. Part. Nucl. Phys.}\ }\textbf {\bibinfo {volume} {93}},\
  \bibinfo {pages} {143} (\bibinfo {year} {2017})},\ \Eprint
  {http://arxiv.org/abs/1610.04528} {arXiv:1610.04528 [hep-ph]} \BibitemShut
  {NoStop}%
\bibitem [{\citenamefont {Guo}\ \emph {et~al.}(2018)\citenamefont {Guo},
  \citenamefont {Hanhart}, \citenamefont {Mei\ss{}ner}, \citenamefont {Wang},
  \citenamefont {Zhao},\ and\ \citenamefont {Zou}}]{Guo:2017jvc}%
  \BibitemOpen
  \bibfield  {author} {\bibinfo {author} {\bibfnamefont {F.-K.}\ \bibnamefont
  {Guo}}, \bibinfo {author} {\bibfnamefont {C.}~\bibnamefont {Hanhart}},
  \bibinfo {author} {\bibfnamefont {U.-G.}\ \bibnamefont {Mei\ss{}ner}},
  \bibinfo {author} {\bibfnamefont {Q.}~\bibnamefont {Wang}}, \bibinfo {author}
  {\bibfnamefont {Q.}~\bibnamefont {Zhao}}, \ and\ \bibinfo {author}
  {\bibfnamefont {B.-S.}\ \bibnamefont {Zou}},\ }\href {\doibase
  10.1103/RevModPhys.90.015004} {\bibfield  {journal} {\bibinfo  {journal}
  {Rev. Mod. Phys.}\ }\textbf {\bibinfo {volume} {90}},\ \bibinfo {pages}
  {015004} (\bibinfo {year} {2018})},\ \bibinfo {note} {[Erratum: Rev. Mod.
  Phys. 94, 029901 (2022)]},\ \Eprint {http://arxiv.org/abs/1705.00141}
  {arXiv:1705.00141 [hep-ph]} \BibitemShut {NoStop}%
\bibitem [{\citenamefont {Olsen}\ \emph {et~al.}(2018)\citenamefont {Olsen},
  \citenamefont {Skwarnicki},\ and\ \citenamefont {Zieminska}}]{Olsen:2017bmm}%
  \BibitemOpen
  \bibfield  {author} {\bibinfo {author} {\bibfnamefont {S.~L.}\ \bibnamefont
  {Olsen}}, \bibinfo {author} {\bibfnamefont {T.}~\bibnamefont {Skwarnicki}}, \
  and\ \bibinfo {author} {\bibfnamefont {D.}~\bibnamefont {Zieminska}},\ }\href
  {\doibase 10.1103/RevModPhys.90.015003} {\bibfield  {journal} {\bibinfo
  {journal} {Rev. Mod. Phys.}\ }\textbf {\bibinfo {volume} {90}},\ \bibinfo
  {pages} {015003} (\bibinfo {year} {2018})},\ \Eprint
  {http://arxiv.org/abs/1708.04012} {arXiv:1708.04012 [hep-ph]} \BibitemShut
  {NoStop}%
\bibitem [{\citenamefont {Liu}\ \emph {et~al.}(2019)\citenamefont {Liu},
  \citenamefont {Chen}, \citenamefont {Chen}, \citenamefont {Liu},\ and\
  \citenamefont {Zhu}}]{Liu:2019zoy}%
  \BibitemOpen
  \bibfield  {author} {\bibinfo {author} {\bibfnamefont {Y.-R.}\ \bibnamefont
  {Liu}}, \bibinfo {author} {\bibfnamefont {H.-X.}\ \bibnamefont {Chen}},
  \bibinfo {author} {\bibfnamefont {W.}~\bibnamefont {Chen}}, \bibinfo {author}
  {\bibfnamefont {X.}~\bibnamefont {Liu}}, \ and\ \bibinfo {author}
  {\bibfnamefont {S.-L.}\ \bibnamefont {Zhu}},\ }\href {\doibase
  10.1016/j.ppnp.2019.04.003} {\bibfield  {journal} {\bibinfo  {journal} {Prog.
  Part. Nucl. Phys.}\ }\textbf {\bibinfo {volume} {107}},\ \bibinfo {pages}
  {237} (\bibinfo {year} {2019})},\ \Eprint {http://arxiv.org/abs/1903.11976}
  {arXiv:1903.11976 [hep-ph]} \BibitemShut {NoStop}%
\bibitem [{\citenamefont {Brambilla}\ \emph {et~al.}(2020)\citenamefont
  {Brambilla}, \citenamefont {Eidelman}, \citenamefont {Hanhart}, \citenamefont
  {Nefediev}, \citenamefont {Shen}, \citenamefont {Thomas}, \citenamefont
  {Vairo},\ and\ \citenamefont {Yuan}}]{Brambilla:2019esw}%
  \BibitemOpen
  \bibfield  {author} {\bibinfo {author} {\bibfnamefont {N.}~\bibnamefont
  {Brambilla}}, \bibinfo {author} {\bibfnamefont {S.}~\bibnamefont {Eidelman}},
  \bibinfo {author} {\bibfnamefont {C.}~\bibnamefont {Hanhart}}, \bibinfo
  {author} {\bibfnamefont {A.}~\bibnamefont {Nefediev}}, \bibinfo {author}
  {\bibfnamefont {C.-P.}\ \bibnamefont {Shen}}, \bibinfo {author}
  {\bibfnamefont {C.~E.}\ \bibnamefont {Thomas}}, \bibinfo {author}
  {\bibfnamefont {A.}~\bibnamefont {Vairo}}, \ and\ \bibinfo {author}
  {\bibfnamefont {C.-Z.}\ \bibnamefont {Yuan}},\ }\href {\doibase
  10.1016/j.physrep.2020.05.001} {\bibfield  {journal} {\bibinfo  {journal}
  {Phys. Rept.}\ }\textbf {\bibinfo {volume} {873}},\ \bibinfo {pages} {1}
  (\bibinfo {year} {2020})},\ \Eprint {http://arxiv.org/abs/1907.07583}
  {arXiv:1907.07583 [hep-ex]} \BibitemShut {NoStop}%
\bibitem [{\citenamefont {Aaij}\ \emph {et~al.}(2015)\citenamefont {Aaij} \emph
  {et~al.}}]{LHCb:2015yax}%
  \BibitemOpen
  \bibfield  {author} {\bibinfo {author} {\bibfnamefont {R.}~\bibnamefont
  {Aaij}} \emph {et~al.} (\bibinfo {collaboration} {LHCb}),\ }\href {\doibase
  10.1103/PhysRevLett.115.072001} {\bibfield  {journal} {\bibinfo  {journal}
  {Phys. Rev. Lett.}\ }\textbf {\bibinfo {volume} {115}},\ \bibinfo {pages}
  {072001} (\bibinfo {year} {2015})},\ \Eprint
  {http://arxiv.org/abs/1507.03414} {arXiv:1507.03414 [hep-ex]} \BibitemShut
  {NoStop}%
\bibitem [{\citenamefont {Wu}\ \emph {et~al.}(2010)\citenamefont {Wu},
  \citenamefont {Molina}, \citenamefont {Oset},\ and\ \citenamefont
  {Zou}}]{Wu:2010jy}%
  \BibitemOpen
  \bibfield  {author} {\bibinfo {author} {\bibfnamefont {J.-J.}\ \bibnamefont
  {Wu}}, \bibinfo {author} {\bibfnamefont {R.}~\bibnamefont {Molina}}, \bibinfo
  {author} {\bibfnamefont {E.}~\bibnamefont {Oset}}, \ and\ \bibinfo {author}
  {\bibfnamefont {B.~S.}\ \bibnamefont {Zou}},\ }\href {\doibase
  10.1103/PhysRevLett.105.232001} {\bibfield  {journal} {\bibinfo  {journal}
  {Phys. Rev. Lett.}\ }\textbf {\bibinfo {volume} {105}},\ \bibinfo {pages}
  {232001} (\bibinfo {year} {2010})},\ \Eprint {http://arxiv.org/abs/1007.0573}
  {arXiv:1007.0573 [nucl-th]} \BibitemShut {NoStop}%
\bibitem [{\citenamefont {Wu}\ \emph {et~al.}(2011)\citenamefont {Wu},
  \citenamefont {Molina}, \citenamefont {Oset},\ and\ \citenamefont
  {Zou}}]{Wu:2010vk}%
  \BibitemOpen
  \bibfield  {author} {\bibinfo {author} {\bibfnamefont {J.-J.}\ \bibnamefont
  {Wu}}, \bibinfo {author} {\bibfnamefont {R.}~\bibnamefont {Molina}}, \bibinfo
  {author} {\bibfnamefont {E.}~\bibnamefont {Oset}}, \ and\ \bibinfo {author}
  {\bibfnamefont {B.~S.}\ \bibnamefont {Zou}},\ }\href {\doibase
  10.1103/PhysRevC.84.015202} {\bibfield  {journal} {\bibinfo  {journal} {Phys.
  Rev. C}\ }\textbf {\bibinfo {volume} {84}},\ \bibinfo {pages} {015202}
  (\bibinfo {year} {2011})},\ \Eprint {http://arxiv.org/abs/1011.2399}
  {arXiv:1011.2399 [nucl-th]} \BibitemShut {NoStop}%
\bibitem [{\citenamefont {Wang}\ \emph {et~al.}(2011)\citenamefont {Wang},
  \citenamefont {Huang}, \citenamefont {Zhang},\ and\ \citenamefont
  {Zou}}]{Wang:2011rga}%
  \BibitemOpen
  \bibfield  {author} {\bibinfo {author} {\bibfnamefont {W.~L.}\ \bibnamefont
  {Wang}}, \bibinfo {author} {\bibfnamefont {F.}~\bibnamefont {Huang}},
  \bibinfo {author} {\bibfnamefont {Z.~Y.}\ \bibnamefont {Zhang}}, \ and\
  \bibinfo {author} {\bibfnamefont {B.~S.}\ \bibnamefont {Zou}},\ }\href
  {\doibase 10.1103/PhysRevC.84.015203} {\bibfield  {journal} {\bibinfo
  {journal} {Phys. Rev. C}\ }\textbf {\bibinfo {volume} {84}},\ \bibinfo
  {pages} {015203} (\bibinfo {year} {2011})},\ \Eprint
  {http://arxiv.org/abs/1101.0453} {arXiv:1101.0453 [nucl-th]} \BibitemShut
  {NoStop}%
\bibitem [{\citenamefont {Yang}\ \emph {et~al.}(2012)\citenamefont {Yang},
  \citenamefont {Sun}, \citenamefont {He}, \citenamefont {Liu},\ and\
  \citenamefont {Zhu}}]{Yang:2011wz}%
  \BibitemOpen
  \bibfield  {author} {\bibinfo {author} {\bibfnamefont {Z.-C.}\ \bibnamefont
  {Yang}}, \bibinfo {author} {\bibfnamefont {Z.-F.}\ \bibnamefont {Sun}},
  \bibinfo {author} {\bibfnamefont {J.}~\bibnamefont {He}}, \bibinfo {author}
  {\bibfnamefont {X.}~\bibnamefont {Liu}}, \ and\ \bibinfo {author}
  {\bibfnamefont {S.-L.}\ \bibnamefont {Zhu}},\ }\href {\doibase
  10.1088/1674-1137/36/1/002} {\bibfield  {journal} {\bibinfo  {journal} {Chin.
  Phys. C}\ }\textbf {\bibinfo {volume} {36}},\ \bibinfo {pages} {6} (\bibinfo
  {year} {2012})},\ \Eprint {http://arxiv.org/abs/1105.2901} {arXiv:1105.2901
  [hep-ph]} \BibitemShut {NoStop}%
\bibitem [{\citenamefont {Oset}\ \emph {et~al.}(2016)\citenamefont {Oset} \emph
  {et~al.}}]{Oset:2016lyh}%
  \BibitemOpen
  \bibfield  {author} {\bibinfo {author} {\bibfnamefont {E.}~\bibnamefont
  {Oset}} \emph {et~al.},\ }\href {\doibase 10.1142/S0218301316300010}
  {\bibfield  {journal} {\bibinfo  {journal} {Int. J. Mod. Phys. E}\ }\textbf
  {\bibinfo {volume} {25}},\ \bibinfo {pages} {1630001} (\bibinfo {year}
  {2016})},\ \Eprint {http://arxiv.org/abs/1601.03972} {arXiv:1601.03972
  [hep-ph]} \BibitemShut {NoStop}%
\bibitem [{\citenamefont {Dong}\ \emph
  {et~al.}(2021{\natexlab{a}})\citenamefont {Dong}, \citenamefont {Guo},\ and\
  \citenamefont {Zou}}]{Dong:2021juy}%
  \BibitemOpen
  \bibfield  {author} {\bibinfo {author} {\bibfnamefont {X.-K.}\ \bibnamefont
  {Dong}}, \bibinfo {author} {\bibfnamefont {F.-K.}\ \bibnamefont {Guo}}, \
  and\ \bibinfo {author} {\bibfnamefont {B.-S.}\ \bibnamefont {Zou}},\ }\href
  {\doibase 10.13725/j.cnki.pip.2021.02.001} {\bibfield  {journal} {\bibinfo
  {journal} {Progr. Phys.}\ }\textbf {\bibinfo {volume} {41}},\ \bibinfo
  {pages} {65} (\bibinfo {year} {2021}{\natexlab{a}})},\ \Eprint
  {http://arxiv.org/abs/2101.01021} {arXiv:2101.01021 [hep-ph]} \BibitemShut
  {NoStop}%
\bibitem [{\citenamefont {Dong}\ \emph
  {et~al.}(2021{\natexlab{b}})\citenamefont {Dong}, \citenamefont {Guo},\ and\
  \citenamefont {Zou}}]{Dong:2021bvy}%
  \BibitemOpen
  \bibfield  {author} {\bibinfo {author} {\bibfnamefont {X.-K.}\ \bibnamefont
  {Dong}}, \bibinfo {author} {\bibfnamefont {F.-K.}\ \bibnamefont {Guo}}, \
  and\ \bibinfo {author} {\bibfnamefont {B.-S.}\ \bibnamefont {Zou}},\ }\href
  {\doibase 10.1088/1572-9494/ac27a2} {\bibfield  {journal} {\bibinfo
  {journal} {Commun. Theor. Phys.}\ }\textbf {\bibinfo {volume} {73}},\
  \bibinfo {pages} {125201} (\bibinfo {year} {2021}{\natexlab{b}})},\ \Eprint
  {http://arxiv.org/abs/2108.02673} {arXiv:2108.02673 [hep-ph]} \BibitemShut
  {NoStop}%
\bibitem [{\citenamefont {Molina}\ \emph {et~al.}(2008)\citenamefont {Molina},
  \citenamefont {Nicmorus},\ and\ \citenamefont {Oset}}]{Molina:2008jw}%
  \BibitemOpen
  \bibfield  {author} {\bibinfo {author} {\bibfnamefont {R.}~\bibnamefont
  {Molina}}, \bibinfo {author} {\bibfnamefont {D.}~\bibnamefont {Nicmorus}}, \
  and\ \bibinfo {author} {\bibfnamefont {E.}~\bibnamefont {Oset}},\ }\href
  {\doibase 10.1103/PhysRevD.78.114018} {\bibfield  {journal} {\bibinfo
  {journal} {Phys. Rev. D}\ }\textbf {\bibinfo {volume} {78}},\ \bibinfo
  {pages} {114018} (\bibinfo {year} {2008})},\ \Eprint
  {http://arxiv.org/abs/0809.2233} {arXiv:0809.2233 [hep-ph]} \BibitemShut
  {NoStop}%
\bibitem [{\citenamefont {Geng}\ and\ \citenamefont
  {Oset}(2009)}]{Geng:2008gx}%
  \BibitemOpen
  \bibfield  {author} {\bibinfo {author} {\bibfnamefont {L.~S.}\ \bibnamefont
  {Geng}}\ and\ \bibinfo {author} {\bibfnamefont {E.}~\bibnamefont {Oset}},\
  }\href {\doibase 10.1103/PhysRevD.79.074009} {\bibfield  {journal} {\bibinfo
  {journal} {Phys. Rev. D}\ }\textbf {\bibinfo {volume} {79}},\ \bibinfo
  {pages} {074009} (\bibinfo {year} {2009})},\ \Eprint
  {http://arxiv.org/abs/0812.1199} {arXiv:0812.1199 [hep-ph]} \BibitemShut
  {NoStop}%
\bibitem [{\citenamefont {Du}\ \emph {et~al.}(2018)\citenamefont {Du},
  \citenamefont {G\"ulmez}, \citenamefont {Guo}, \citenamefont {Mei\ss{}ner},\
  and\ \citenamefont {Wang}}]{Du:2018gyn}%
  \BibitemOpen
  \bibfield  {author} {\bibinfo {author} {\bibfnamefont {M.-L.}\ \bibnamefont
  {Du}}, \bibinfo {author} {\bibfnamefont {D.}~\bibnamefont {G\"ulmez}},
  \bibinfo {author} {\bibfnamefont {F.-K.}\ \bibnamefont {Guo}}, \bibinfo
  {author} {\bibfnamefont {U.-G.}\ \bibnamefont {Mei\ss{}ner}}, \ and\ \bibinfo
  {author} {\bibfnamefont {Q.}~\bibnamefont {Wang}},\ }\href {\doibase
  10.1140/epjc/s10052-018-6475-8} {\bibfield  {journal} {\bibinfo  {journal}
  {Eur. Phys. J. C}\ }\textbf {\bibinfo {volume} {78}},\ \bibinfo {pages} {988}
  (\bibinfo {year} {2018})},\ \Eprint {http://arxiv.org/abs/1808.09664}
  {arXiv:1808.09664 [hep-ph]} \BibitemShut {NoStop}%
\bibitem [{\citenamefont {Nagahiro}\ \emph {et~al.}(2008)\citenamefont
  {Nagahiro}, \citenamefont {Roca}, \citenamefont {Oset},\ and\ \citenamefont
  {Zou}}]{Nagahiro:2008bn}%
  \BibitemOpen
  \bibfield  {author} {\bibinfo {author} {\bibfnamefont {H.}~\bibnamefont
  {Nagahiro}}, \bibinfo {author} {\bibfnamefont {L.}~\bibnamefont {Roca}},
  \bibinfo {author} {\bibfnamefont {E.}~\bibnamefont {Oset}}, \ and\ \bibinfo
  {author} {\bibfnamefont {B.~S.}\ \bibnamefont {Zou}},\ }\href {\doibase
  10.1103/PhysRevD.78.014012} {\bibfield  {journal} {\bibinfo  {journal} {Phys.
  Rev. D}\ }\textbf {\bibinfo {volume} {78}},\ \bibinfo {pages} {014012}
  (\bibinfo {year} {2008})},\ \Eprint {http://arxiv.org/abs/0803.4460}
  {arXiv:0803.4460 [hep-ph]} \BibitemShut {NoStop}%
\bibitem [{\citenamefont {Branz}\ \emph {et~al.}(2010)\citenamefont {Branz},
  \citenamefont {Geng},\ and\ \citenamefont {Oset}}]{Branz:2009cv}%
  \BibitemOpen
  \bibfield  {author} {\bibinfo {author} {\bibfnamefont {T.}~\bibnamefont
  {Branz}}, \bibinfo {author} {\bibfnamefont {L.~S.}\ \bibnamefont {Geng}}, \
  and\ \bibinfo {author} {\bibfnamefont {E.}~\bibnamefont {Oset}},\ }\href
  {\doibase 10.1103/PhysRevD.81.054037} {\bibfield  {journal} {\bibinfo
  {journal} {Phys. Rev. D}\ }\textbf {\bibinfo {volume} {81}},\ \bibinfo
  {pages} {054037} (\bibinfo {year} {2010})},\ \Eprint
  {http://arxiv.org/abs/0911.0206} {arXiv:0911.0206 [hep-ph]} \BibitemShut
  {NoStop}%
\bibitem [{\citenamefont {Geng}\ \emph {et~al.}(2009)\citenamefont {Geng},
  \citenamefont {Oset}, \citenamefont {Molina},\ and\ \citenamefont
  {Nicmorus}}]{Geng:2009gb}%
  \BibitemOpen
  \bibfield  {author} {\bibinfo {author} {\bibfnamefont {L.~S.}\ \bibnamefont
  {Geng}}, \bibinfo {author} {\bibfnamefont {E.}~\bibnamefont {Oset}}, \bibinfo
  {author} {\bibfnamefont {R.}~\bibnamefont {Molina}}, \ and\ \bibinfo {author}
  {\bibfnamefont {D.}~\bibnamefont {Nicmorus}},\ }\href {\doibase
  10.22323/1.069.0040} {\bibfield  {journal} {\bibinfo  {journal} {PoS}\
  }\textbf {\bibinfo {volume} {EFT09}},\ \bibinfo {pages} {040} (\bibinfo
  {year} {2009})},\ \Eprint {http://arxiv.org/abs/0905.0419} {arXiv:0905.0419
  [hep-ph]} \BibitemShut {NoStop}%
\bibitem [{\citenamefont {Geng}\ \emph {et~al.}(2010)\citenamefont {Geng},
  \citenamefont {Guo}, \citenamefont {Hanhart}, \citenamefont {Molina},
  \citenamefont {Oset},\ and\ \citenamefont {Zou}}]{Geng:2010kma}%
  \BibitemOpen
  \bibfield  {author} {\bibinfo {author} {\bibfnamefont {L.~S.}\ \bibnamefont
  {Geng}}, \bibinfo {author} {\bibfnamefont {F.~K.}\ \bibnamefont {Guo}},
  \bibinfo {author} {\bibfnamefont {C.}~\bibnamefont {Hanhart}}, \bibinfo
  {author} {\bibfnamefont {R.}~\bibnamefont {Molina}}, \bibinfo {author}
  {\bibfnamefont {E.}~\bibnamefont {Oset}}, \ and\ \bibinfo {author}
  {\bibfnamefont {B.~S.}\ \bibnamefont {Zou}},\ }\href {\doibase
  10.1140/epja/i2010-10971-5} {\bibfield  {journal} {\bibinfo  {journal} {Eur.
  Phys. J. A}\ }\textbf {\bibinfo {volume} {44}},\ \bibinfo {pages} {305}
  (\bibinfo {year} {2010})},\ \Eprint {http://arxiv.org/abs/0910.5192}
  {arXiv:0910.5192 [hep-ph]} \BibitemShut {NoStop}%
\bibitem [{\citenamefont {Martinez~Torres}\ \emph {et~al.}(2013)\citenamefont
  {Martinez~Torres}, \citenamefont {Khemchandani}, \citenamefont {Navarra},
  \citenamefont {Nielsen},\ and\ \citenamefont {Oset}}]{MartinezTorres:2012du}%
  \BibitemOpen
  \bibfield  {author} {\bibinfo {author} {\bibfnamefont {A.}~\bibnamefont
  {Martinez~Torres}}, \bibinfo {author} {\bibfnamefont {K.~P.}\ \bibnamefont
  {Khemchandani}}, \bibinfo {author} {\bibfnamefont {F.~S.}\ \bibnamefont
  {Navarra}}, \bibinfo {author} {\bibfnamefont {M.}~\bibnamefont {Nielsen}}, \
  and\ \bibinfo {author} {\bibfnamefont {E.}~\bibnamefont {Oset}},\ }\href
  {\doibase 10.1016/j.physletb.2013.01.036} {\bibfield  {journal} {\bibinfo
  {journal} {Phys. Lett. B}\ }\textbf {\bibinfo {volume} {719}},\ \bibinfo
  {pages} {388} (\bibinfo {year} {2013})},\ \Eprint
  {http://arxiv.org/abs/1210.6392} {arXiv:1210.6392 [hep-ph]} \BibitemShut
  {NoStop}%
\bibitem [{\citenamefont {Xie}\ and\ \citenamefont {Oset}(2014)}]{Xie:2014gla}%
  \BibitemOpen
  \bibfield  {author} {\bibinfo {author} {\bibfnamefont {J.-J.}\ \bibnamefont
  {Xie}}\ and\ \bibinfo {author} {\bibfnamefont {E.}~\bibnamefont {Oset}},\
  }\href {\doibase 10.1103/PhysRevD.90.094006} {\bibfield  {journal} {\bibinfo
  {journal} {Phys. Rev. D}\ }\textbf {\bibinfo {volume} {90}},\ \bibinfo
  {pages} {094006} (\bibinfo {year} {2014})},\ \Eprint
  {http://arxiv.org/abs/1409.1341} {arXiv:1409.1341 [hep-ph]} \BibitemShut
  {NoStop}%
\bibitem [{\citenamefont {Molina}\ \emph {et~al.}(2020)\citenamefont {Molina},
  \citenamefont {Dai}, \citenamefont {Geng},\ and\ \citenamefont
  {Oset}}]{Molina:2019wjj}%
  \BibitemOpen
  \bibfield  {author} {\bibinfo {author} {\bibfnamefont {R.}~\bibnamefont
  {Molina}}, \bibinfo {author} {\bibfnamefont {L.~R.}\ \bibnamefont {Dai}},
  \bibinfo {author} {\bibfnamefont {L.~S.}\ \bibnamefont {Geng}}, \ and\
  \bibinfo {author} {\bibfnamefont {E.}~\bibnamefont {Oset}},\ }\href {\doibase
  10.1140/epja/s10050-020-00176-y} {\bibfield  {journal} {\bibinfo  {journal}
  {Eur. Phys. J. A}\ }\textbf {\bibinfo {volume} {56}},\ \bibinfo {pages} {173}
  (\bibinfo {year} {2020})},\ \Eprint {http://arxiv.org/abs/1909.10764}
  {arXiv:1909.10764 [hep-ph]} \BibitemShut {NoStop}%
\bibitem [{\citenamefont {Wang}\ and\ \citenamefont
  {Zou}(2021)}]{Wang:2021jub}%
  \BibitemOpen
  \bibfield  {author} {\bibinfo {author} {\bibfnamefont {Z.-L.}\ \bibnamefont
  {Wang}}\ and\ \bibinfo {author} {\bibfnamefont {B.-S.}\ \bibnamefont {Zou}},\
  }\href {\doibase 10.1103/PhysRevD.104.114001} {\bibfield  {journal} {\bibinfo
   {journal} {Phys. Rev. D}\ }\textbf {\bibinfo {volume} {104}},\ \bibinfo
  {pages} {114001} (\bibinfo {year} {2021})},\ \Eprint
  {http://arxiv.org/abs/2107.14470} {arXiv:2107.14470 [hep-ph]} \BibitemShut
  {NoStop}%
\bibitem [{\citenamefont {Wang}\ and\ \citenamefont
  {Zou}(2022)}]{Wang:2022pin}%
  \BibitemOpen
  \bibfield  {author} {\bibinfo {author} {\bibfnamefont {Z.-L.}\ \bibnamefont
  {Wang}}\ and\ \bibinfo {author} {\bibfnamefont {B.-S.}\ \bibnamefont {Zou}},\
  }\href {\doibase 10.1140/epjc/s10052-022-10460-4} {\bibfield  {journal}
  {\bibinfo  {journal} {Eur. Phys. J. C}\ }\textbf {\bibinfo {volume} {82}},\
  \bibinfo {pages} {509} (\bibinfo {year} {2022})},\ \Eprint
  {http://arxiv.org/abs/2203.02899} {arXiv:2203.02899 [hep-ph]} \BibitemShut
  {NoStop}%
\bibitem [{\citenamefont {Lees}\ \emph {et~al.}(2021)\citenamefont {Lees} \emph
  {et~al.}}]{BaBar:2021fkz}%
  \BibitemOpen
  \bibfield  {author} {\bibinfo {author} {\bibfnamefont {J.~P.}\ \bibnamefont
  {Lees}} \emph {et~al.} (\bibinfo {collaboration} {BaBar}),\ }\href {\doibase
  10.1103/PhysRevD.104.072002} {\bibfield  {journal} {\bibinfo  {journal}
  {Phys. Rev. D}\ }\textbf {\bibinfo {volume} {104}},\ \bibinfo {pages}
  {072002} (\bibinfo {year} {2021})},\ \Eprint
  {http://arxiv.org/abs/2106.05157} {arXiv:2106.05157 [hep-ex]} \BibitemShut
  {NoStop}%
\bibitem [{\citenamefont {Ablikim}\ \emph {et~al.}(2022)\citenamefont {Ablikim}
  \emph {et~al.}}]{BESIII:2022npc}%
  \BibitemOpen
  \bibfield  {author} {\bibinfo {author} {\bibfnamefont {M.}~\bibnamefont
  {Ablikim}} \emph {et~al.} (\bibinfo {collaboration} {BESIII}),\ }\href
  {\doibase 10.1103/PhysRevLett.129.182001} {\bibfield  {journal} {\bibinfo
  {journal} {Phys. Rev. Lett.}\ }\textbf {\bibinfo {volume} {129}},\ \bibinfo
  {pages} {182001} (\bibinfo {year} {2022})},\ \Eprint
  {http://arxiv.org/abs/2204.09614} {arXiv:2204.09614 [hep-ex]} \BibitemShut
  {NoStop}%
\bibitem [{\citenamefont {Dai}\ \emph {et~al.}(2022)\citenamefont {Dai},
  \citenamefont {Oset},\ and\ \citenamefont {Geng}}]{Dai:2021owu}%
  \BibitemOpen
  \bibfield  {author} {\bibinfo {author} {\bibfnamefont {L.~R.}\ \bibnamefont
  {Dai}}, \bibinfo {author} {\bibfnamefont {E.}~\bibnamefont {Oset}}, \ and\
  \bibinfo {author} {\bibfnamefont {L.~S.}\ \bibnamefont {Geng}},\ }\href
  {\doibase 10.1140/epjc/s10052-022-10178-3} {\bibfield  {journal} {\bibinfo
  {journal} {Eur. Phys. J. C}\ }\textbf {\bibinfo {volume} {82}},\ \bibinfo
  {pages} {225} (\bibinfo {year} {2022})},\ \Eprint
  {http://arxiv.org/abs/2111.10230} {arXiv:2111.10230 [hep-ph]} \BibitemShut
  {NoStop}%
\bibitem [{\citenamefont {Zhu}\ \emph {et~al.}(2022)\citenamefont {Zhu},
  \citenamefont {Li}, \citenamefont {Wang}, \citenamefont {Geng},\ and\
  \citenamefont {Xie}}]{Zhu:2022wzk}%
  \BibitemOpen
  \bibfield  {author} {\bibinfo {author} {\bibfnamefont {X.}~\bibnamefont
  {Zhu}}, \bibinfo {author} {\bibfnamefont {D.-M.}\ \bibnamefont {Li}},
  \bibinfo {author} {\bibfnamefont {E.}~\bibnamefont {Wang}}, \bibinfo {author}
  {\bibfnamefont {L.-S.}\ \bibnamefont {Geng}}, \ and\ \bibinfo {author}
  {\bibfnamefont {J.-J.}\ \bibnamefont {Xie}},\ }\href {\doibase
  10.1103/PhysRevD.105.116010} {\bibfield  {journal} {\bibinfo  {journal}
  {Phys. Rev. D}\ }\textbf {\bibinfo {volume} {105}},\ \bibinfo {pages}
  {116010} (\bibinfo {year} {2022})},\ \Eprint
  {http://arxiv.org/abs/2204.09384} {arXiv:2204.09384 [hep-ph]} \BibitemShut
  {NoStop}%
\bibitem [{\citenamefont {Zhu}\ \emph {et~al.}(2023)\citenamefont {Zhu},
  \citenamefont {Wang}, \citenamefont {Li}, \citenamefont {Wang}, \citenamefont
  {Geng},\ and\ \citenamefont {Xie}}]{Zhu:2022guw}%
  \BibitemOpen
  \bibfield  {author} {\bibinfo {author} {\bibfnamefont {X.}~\bibnamefont
  {Zhu}}, \bibinfo {author} {\bibfnamefont {H.-N.}\ \bibnamefont {Wang}},
  \bibinfo {author} {\bibfnamefont {D.-M.}\ \bibnamefont {Li}}, \bibinfo
  {author} {\bibfnamefont {E.}~\bibnamefont {Wang}}, \bibinfo {author}
  {\bibfnamefont {L.-S.}\ \bibnamefont {Geng}}, \ and\ \bibinfo {author}
  {\bibfnamefont {J.-J.}\ \bibnamefont {Xie}},\ }\href {\doibase
  10.1103/PhysRevD.107.034001} {\bibfield  {journal} {\bibinfo  {journal}
  {Phys.Rev.D}\ }\textbf {\bibinfo {volume} {107}},\ \bibinfo {pages} {034001}
  (\bibinfo {year} {2023})},\ \Eprint {http://arxiv.org/abs/2210.12992}
  {arXiv:2210.12992 [hep-ph]} \BibitemShut {NoStop}%
\bibitem [{\citenamefont {Martinez~Torres}\ \emph
  {et~al.}(2011{\natexlab{a}})\citenamefont {Martinez~Torres}, \citenamefont
  {Khemchandani}, \citenamefont {Jido},\ and\ \citenamefont
  {Hosaka}}]{MartinezTorres:2011vh}%
  \BibitemOpen
  \bibfield  {author} {\bibinfo {author} {\bibfnamefont {A.}~\bibnamefont
  {Martinez~Torres}}, \bibinfo {author} {\bibfnamefont {K.~P.}\ \bibnamefont
  {Khemchandani}}, \bibinfo {author} {\bibfnamefont {D.}~\bibnamefont {Jido}},
  \ and\ \bibinfo {author} {\bibfnamefont {A.}~\bibnamefont {Hosaka}},\ }\href
  {\doibase 10.1103/PhysRevD.84.074027} {\bibfield  {journal} {\bibinfo
  {journal} {Phys. Rev. D}\ }\textbf {\bibinfo {volume} {84}},\ \bibinfo
  {pages} {074027} (\bibinfo {year} {2011}{\natexlab{a}})},\ \Eprint
  {http://arxiv.org/abs/1106.6101} {arXiv:1106.6101 [nucl-th]} \BibitemShut
  {NoStop}%
\bibitem [{\citenamefont {Martinez~Torres}\ \emph
  {et~al.}(2011{\natexlab{b}})\citenamefont {Martinez~Torres}, \citenamefont
  {Jido},\ and\ \citenamefont {Kanada-En'yo}}]{MartinezTorres:2011gjk}%
  \BibitemOpen
  \bibfield  {author} {\bibinfo {author} {\bibfnamefont {A.}~\bibnamefont
  {Martinez~Torres}}, \bibinfo {author} {\bibfnamefont {D.}~\bibnamefont
  {Jido}}, \ and\ \bibinfo {author} {\bibfnamefont {Y.}~\bibnamefont
  {Kanada-En'yo}},\ }\href {\doibase 10.1103/PhysRevC.83.065205} {\bibfield
  {journal} {\bibinfo  {journal} {Phys. Rev. C}\ }\textbf {\bibinfo {volume}
  {83}},\ \bibinfo {pages} {065205} (\bibinfo {year} {2011}{\natexlab{b}})},\
  \Eprint {http://arxiv.org/abs/1102.1505} {arXiv:1102.1505 [nucl-th]}
  \BibitemShut {NoStop}%
\bibitem [{\citenamefont {Liang}\ \emph {et~al.}(2013)\citenamefont {Liang},
  \citenamefont {Xiao},\ and\ \citenamefont {Oset}}]{Liang:2013yta}%
  \BibitemOpen
  \bibfield  {author} {\bibinfo {author} {\bibfnamefont {W.}~\bibnamefont
  {Liang}}, \bibinfo {author} {\bibfnamefont {C.~W.}\ \bibnamefont {Xiao}}, \
  and\ \bibinfo {author} {\bibfnamefont {E.}~\bibnamefont {Oset}},\ }\href
  {\doibase 10.1103/PhysRevD.88.114024} {\bibfield  {journal} {\bibinfo
  {journal} {Phys. Rev. D}\ }\textbf {\bibinfo {volume} {88}},\ \bibinfo
  {pages} {114024} (\bibinfo {year} {2013})},\ \Eprint
  {http://arxiv.org/abs/1309.7310} {arXiv:1309.7310 [hep-ph]} \BibitemShut
  {NoStop}%
\bibitem [{\citenamefont {Zhang}\ \emph {et~al.}(2017)\citenamefont {Zhang},
  \citenamefont {Xie},\ and\ \citenamefont {Chen}}]{Zhang:2016bmy}%
  \BibitemOpen
  \bibfield  {author} {\bibinfo {author} {\bibfnamefont {X.}~\bibnamefont
  {Zhang}}, \bibinfo {author} {\bibfnamefont {J.-J.}\ \bibnamefont {Xie}}, \
  and\ \bibinfo {author} {\bibfnamefont {X.}~\bibnamefont {Chen}},\ }\href
  {\doibase 10.1103/PhysRevD.95.056014} {\bibfield  {journal} {\bibinfo
  {journal} {Phys. Rev. D}\ }\textbf {\bibinfo {volume} {95}},\ \bibinfo
  {pages} {056014} (\bibinfo {year} {2017})},\ \Eprint
  {http://arxiv.org/abs/1612.02613} {arXiv:1612.02613 [hep-ph]} \BibitemShut
  {NoStop}%
\bibitem [{\citenamefont {Bayar}\ \emph {et~al.}(2014)\citenamefont {Bayar},
  \citenamefont {Liang}, \citenamefont {Uchino},\ and\ \citenamefont
  {Xiao}}]{Bayar:2013bta}%
  \BibitemOpen
  \bibfield  {author} {\bibinfo {author} {\bibfnamefont {M.}~\bibnamefont
  {Bayar}}, \bibinfo {author} {\bibfnamefont {W.~H.}\ \bibnamefont {Liang}},
  \bibinfo {author} {\bibfnamefont {T.}~\bibnamefont {Uchino}}, \ and\ \bibinfo
  {author} {\bibfnamefont {C.~W.}\ \bibnamefont {Xiao}},\ }\href {\doibase
  10.1140/epja/i2014-14067-0} {\bibfield  {journal} {\bibinfo  {journal} {Eur.
  Phys. J. A}\ }\textbf {\bibinfo {volume} {50}},\ \bibinfo {pages} {67}
  (\bibinfo {year} {2014})},\ \Eprint {http://arxiv.org/abs/1312.2869}
  {arXiv:1312.2869 [hep-ph]} \BibitemShut {NoStop}%
\bibitem [{\citenamefont {Martinez~Torres}\ \emph
  {et~al.}(2008{\natexlab{a}})\citenamefont {Martinez~Torres}, \citenamefont
  {Khemchandani}, \citenamefont {Geng}, \citenamefont {Napsuciale},\ and\
  \citenamefont {Oset}}]{MartinezTorres:2008gy}%
  \BibitemOpen
  \bibfield  {author} {\bibinfo {author} {\bibfnamefont {A.}~\bibnamefont
  {Martinez~Torres}}, \bibinfo {author} {\bibfnamefont {K.~P.}\ \bibnamefont
  {Khemchandani}}, \bibinfo {author} {\bibfnamefont {L.~S.}\ \bibnamefont
  {Geng}}, \bibinfo {author} {\bibfnamefont {M.}~\bibnamefont {Napsuciale}}, \
  and\ \bibinfo {author} {\bibfnamefont {E.}~\bibnamefont {Oset}},\ }\href
  {\doibase 10.1103/PhysRevD.78.074031} {\bibfield  {journal} {\bibinfo
  {journal} {Phys. Rev. D}\ }\textbf {\bibinfo {volume} {78}},\ \bibinfo
  {pages} {074031} (\bibinfo {year} {2008}{\natexlab{a}})},\ \Eprint
  {http://arxiv.org/abs/0801.3635} {arXiv:0801.3635 [nucl-th]} \BibitemShut
  {NoStop}%
\bibitem [{\citenamefont {Martinez~Torres}\ \emph
  {et~al.}(2009{\natexlab{a}})\citenamefont {Martinez~Torres}, \citenamefont
  {Khemchandani}, \citenamefont {Gamermann},\ and\ \citenamefont
  {Oset}}]{MartinezTorres:2009xb}%
  \BibitemOpen
  \bibfield  {author} {\bibinfo {author} {\bibfnamefont {A.}~\bibnamefont
  {Martinez~Torres}}, \bibinfo {author} {\bibfnamefont {K.~P.}\ \bibnamefont
  {Khemchandani}}, \bibinfo {author} {\bibfnamefont {D.}~\bibnamefont
  {Gamermann}}, \ and\ \bibinfo {author} {\bibfnamefont {E.}~\bibnamefont
  {Oset}},\ }\href {\doibase 10.1103/PhysRevD.80.094012} {\bibfield  {journal}
  {\bibinfo  {journal} {Phys. Rev. D}\ }\textbf {\bibinfo {volume} {80}},\
  \bibinfo {pages} {094012} (\bibinfo {year} {2009}{\natexlab{a}})},\ \Eprint
  {http://arxiv.org/abs/0906.5333} {arXiv:0906.5333 [nucl-th]} \BibitemShut
  {NoStop}%
\bibitem [{\citenamefont {Martinez~Torres}\ \emph {et~al.}(2020)\citenamefont
  {Martinez~Torres}, \citenamefont {Khemchandani}, \citenamefont {Roca},\ and\
  \citenamefont {Oset}}]{MartinezTorres:2020hus}%
  \BibitemOpen
  \bibfield  {author} {\bibinfo {author} {\bibfnamefont {A.}~\bibnamefont
  {Martinez~Torres}}, \bibinfo {author} {\bibfnamefont {K.~P.}\ \bibnamefont
  {Khemchandani}}, \bibinfo {author} {\bibfnamefont {L.}~\bibnamefont {Roca}},
  \ and\ \bibinfo {author} {\bibfnamefont {E.}~\bibnamefont {Oset}},\ }\href
  {\doibase 10.1007/s00601-020-01568-y} {\bibfield  {journal} {\bibinfo
  {journal} {Few Body Syst.}\ }\textbf {\bibinfo {volume} {61}},\ \bibinfo
  {pages} {35} (\bibinfo {year} {2020})},\ \Eprint
  {http://arxiv.org/abs/2005.14357} {arXiv:2005.14357 [nucl-th]} \BibitemShut
  {NoStop}%
\bibitem [{\citenamefont {Wu}\ \emph {et~al.}(2022)\citenamefont {Wu},
  \citenamefont {Pan}, \citenamefont {Liu},\ and\ \citenamefont
  {Geng}}]{Wu:2022ftm}%
  \BibitemOpen
  \bibfield  {author} {\bibinfo {author} {\bibfnamefont {T.-W.}\ \bibnamefont
  {Wu}}, \bibinfo {author} {\bibfnamefont {Y.-W.}\ \bibnamefont {Pan}},
  \bibinfo {author} {\bibfnamefont {M.-Z.}\ \bibnamefont {Liu}}, \ and\
  \bibinfo {author} {\bibfnamefont {L.-S.}\ \bibnamefont {Geng}},\ }\href
  {\doibase 10.1016/j.scib.2022.08.007} {\bibfield  {journal} {\bibinfo
  {journal} {Sci. Bull.}\ }\textbf {\bibinfo {volume} {67}},\ \bibinfo {pages}
  {1735} (\bibinfo {year} {2022})},\ \Eprint {http://arxiv.org/abs/2208.00882}
  {arXiv:2208.00882 [hep-ph]} \BibitemShut {NoStop}%
\bibitem [{\citenamefont {Malabarba}\ \emph {et~al.}(2022)\citenamefont
  {Malabarba}, \citenamefont {Khemchandani},\ and\ \citenamefont
  {Torres}}]{Malabarba:2021taj}%
  \BibitemOpen
  \bibfield  {author} {\bibinfo {author} {\bibfnamefont {B.~B.}\ \bibnamefont
  {Malabarba}}, \bibinfo {author} {\bibfnamefont {K.~P.}\ \bibnamefont
  {Khemchandani}}, \ and\ \bibinfo {author} {\bibfnamefont {A.~M.}\
  \bibnamefont {Torres}},\ }\href {\doibase 10.1140/epja/s10050-022-00681-2}
  {\bibfield  {journal} {\bibinfo  {journal} {Eur. Phys. J. A}\ }\textbf
  {\bibinfo {volume} {58}},\ \bibinfo {pages} {33} (\bibinfo {year} {2022})},\
  \Eprint {http://arxiv.org/abs/2103.09978} {arXiv:2103.09978 [hep-ph]}
  \BibitemShut {NoStop}%
\bibitem [{\citenamefont {Debastiani}\ \emph {et~al.}(2017)\citenamefont
  {Debastiani}, \citenamefont {Dias},\ and\ \citenamefont
  {Oset}}]{Debastiani:2017vhv}%
  \BibitemOpen
  \bibfield  {author} {\bibinfo {author} {\bibfnamefont {V.~R.}\ \bibnamefont
  {Debastiani}}, \bibinfo {author} {\bibfnamefont {J.~M.}\ \bibnamefont
  {Dias}}, \ and\ \bibinfo {author} {\bibfnamefont {E.}~\bibnamefont {Oset}},\
  }\href {\doibase 10.1103/PhysRevD.96.016014} {\bibfield  {journal} {\bibinfo
  {journal} {Phys. Rev. D}\ }\textbf {\bibinfo {volume} {96}},\ \bibinfo
  {pages} {016014} (\bibinfo {year} {2017})},\ \Eprint
  {http://arxiv.org/abs/1705.09257} {arXiv:1705.09257 [hep-ph]} \BibitemShut
  {NoStop}%
\bibitem [{\citenamefont {Malabarba}\ \emph {et~al.}(2023)\citenamefont
  {Malabarba}, \citenamefont {Khemchandani}, \citenamefont {Martinez~Torres},\
  and\ \citenamefont {Oset}}]{Malabarba:2022pdo}%
  \BibitemOpen
  \bibfield  {author} {\bibinfo {author} {\bibfnamefont {B.~B.}\ \bibnamefont
  {Malabarba}}, \bibinfo {author} {\bibfnamefont {K.~P.}\ \bibnamefont
  {Khemchandani}}, \bibinfo {author} {\bibfnamefont {A.}~\bibnamefont
  {Martinez~Torres}}, \ and\ \bibinfo {author} {\bibfnamefont {E.}~\bibnamefont
  {Oset}},\ }\href {\doibase 10.1103/PhysRevD.107.036016} {\bibfield  {journal}
  {\bibinfo  {journal} {Phys. Rev. D}\ }\textbf {\bibinfo {volume} {107}},\
  \bibinfo {pages} {036016} (\bibinfo {year} {2023})},\ \Eprint
  {http://arxiv.org/abs/2211.16222} {arXiv:2211.16222 [hep-ph]} \BibitemShut
  {NoStop}%
\bibitem [{\citenamefont {Luo}\ \emph {et~al.}(2022{\natexlab{a}})\citenamefont
  {Luo}, \citenamefont {Wu}, \citenamefont {Liu}, \citenamefont {Geng},\ and\
  \citenamefont {Liu}}]{Luo:2021ggs}%
  \BibitemOpen
  \bibfield  {author} {\bibinfo {author} {\bibfnamefont {S.-Q.}\ \bibnamefont
  {Luo}}, \bibinfo {author} {\bibfnamefont {T.-W.}\ \bibnamefont {Wu}},
  \bibinfo {author} {\bibfnamefont {M.-Z.}\ \bibnamefont {Liu}}, \bibinfo
  {author} {\bibfnamefont {L.-S.}\ \bibnamefont {Geng}}, \ and\ \bibinfo
  {author} {\bibfnamefont {X.}~\bibnamefont {Liu}},\ }\href {\doibase
  10.1103/PhysRevD.105.074033} {\bibfield  {journal} {\bibinfo  {journal}
  {Phys. Rev. D}\ }\textbf {\bibinfo {volume} {105}},\ \bibinfo {pages}
  {074033} (\bibinfo {year} {2022}{\natexlab{a}})},\ \Eprint
  {http://arxiv.org/abs/2111.15079} {arXiv:2111.15079 [hep-ph]} \BibitemShut
  {NoStop}%
\bibitem [{\citenamefont {Luo}\ \emph {et~al.}(2022{\natexlab{b}})\citenamefont
  {Luo}, \citenamefont {Geng},\ and\ \citenamefont {Liu}}]{Luo:2022cun}%
  \BibitemOpen
  \bibfield  {author} {\bibinfo {author} {\bibfnamefont {S.-Q.}\ \bibnamefont
  {Luo}}, \bibinfo {author} {\bibfnamefont {L.-S.}\ \bibnamefont {Geng}}, \
  and\ \bibinfo {author} {\bibfnamefont {X.}~\bibnamefont {Liu}},\ }\href
  {\doibase 10.1103/PhysRevD.106.014017} {\bibfield  {journal} {\bibinfo
  {journal} {Phys. Rev. D}\ }\textbf {\bibinfo {volume} {106}},\ \bibinfo
  {pages} {014017} (\bibinfo {year} {2022}{\natexlab{b}})},\ \Eprint
  {http://arxiv.org/abs/2206.04586} {arXiv:2206.04586 [hep-ph]} \BibitemShut
  {NoStop}%
\bibitem [{\citenamefont {Ikeno}\ \emph {et~al.}(2023)\citenamefont {Ikeno},
  \citenamefont {Bayar},\ and\ \citenamefont {Oset}}]{Ikeno:2022jbb}%
  \BibitemOpen
  \bibfield  {author} {\bibinfo {author} {\bibfnamefont {N.}~\bibnamefont
  {Ikeno}}, \bibinfo {author} {\bibfnamefont {M.}~\bibnamefont {Bayar}}, \ and\
  \bibinfo {author} {\bibfnamefont {E.}~\bibnamefont {Oset}},\ }\href {\doibase
  10.1103/PhysRevD.107.034006} {\bibfield  {journal} {\bibinfo  {journal}
  {Phys. Rev. D}\ }\textbf {\bibinfo {volume} {107}},\ \bibinfo {pages}
  {034006} (\bibinfo {year} {2023})},\ \Eprint
  {http://arxiv.org/abs/2208.03698} {arXiv:2208.03698 [hep-ph]} \BibitemShut
  {NoStop}%
\bibitem [{\citenamefont {Wu}\ \emph {et~al.}(2021)\citenamefont {Wu},
  \citenamefont {Liu},\ and\ \citenamefont {Geng}}]{Wu:2020job}%
  \BibitemOpen
  \bibfield  {author} {\bibinfo {author} {\bibfnamefont {T.-W.}\ \bibnamefont
  {Wu}}, \bibinfo {author} {\bibfnamefont {M.-Z.}\ \bibnamefont {Liu}}, \ and\
  \bibinfo {author} {\bibfnamefont {L.-S.}\ \bibnamefont {Geng}},\ }\href
  {\doibase 10.1103/PhysRevD.103.L031501} {\bibfield  {journal} {\bibinfo
  {journal} {Phys. Rev. D}\ }\textbf {\bibinfo {volume} {103}},\ \bibinfo
  {pages} {L031501} (\bibinfo {year} {2021})},\ \Eprint
  {http://arxiv.org/abs/2012.01134} {arXiv:2012.01134 [hep-ph]} \BibitemShut
  {NoStop}%
\bibitem [{\citenamefont {Wei}\ \emph {et~al.}(2022)\citenamefont {Wei},
  \citenamefont {Shen},\ and\ \citenamefont {Xie}}]{Wei:2022jgc}%
  \BibitemOpen
  \bibfield  {author} {\bibinfo {author} {\bibfnamefont {X.}~\bibnamefont
  {Wei}}, \bibinfo {author} {\bibfnamefont {Q.-H.}\ \bibnamefont {Shen}}, \
  and\ \bibinfo {author} {\bibfnamefont {J.-J.}\ \bibnamefont {Xie}},\ }\href
  {\doibase 10.1140/epjc/s10052-022-10675-5} {\bibfield  {journal} {\bibinfo
  {journal} {Eur. Phys. J. C}\ }\textbf {\bibinfo {volume} {82}},\ \bibinfo
  {pages} {718} (\bibinfo {year} {2022})},\ \Eprint
  {http://arxiv.org/abs/2205.12526} {arXiv:2205.12526 [hep-ph]} \BibitemShut
  {NoStop}%
\bibitem [{\citenamefont {Bayar}\ \emph {et~al.}(2023)\citenamefont {Bayar},
  \citenamefont {Ikeno},\ and\ \citenamefont {Roca}}]{Bayar:2023itf}%
  \BibitemOpen
  \bibfield  {author} {\bibinfo {author} {\bibfnamefont {M.}~\bibnamefont
  {Bayar}}, \bibinfo {author} {\bibfnamefont {N.}~\bibnamefont {Ikeno}}, \ and\
  \bibinfo {author} {\bibfnamefont {L.}~\bibnamefont {Roca}},\ }\href {\doibase
  10.1103/PhysRevD.107.054042} {\bibfield  {journal} {\bibinfo  {journal}
  {Phys. Rev. D}\ }\textbf {\bibinfo {volume} {107}},\ \bibinfo {pages}
  {054042} (\bibinfo {year} {2023})},\ \Eprint
  {http://arxiv.org/abs/2301.07436} {arXiv:2301.07436 [hep-ph]} \BibitemShut
  {NoStop}%
\bibitem [{\citenamefont {Hiyama}\ \emph {et~al.}(2003)\citenamefont {Hiyama},
  \citenamefont {Kino},\ and\ \citenamefont {Kamimura}}]{Hiyama:2003cu}%
  \BibitemOpen
  \bibfield  {author} {\bibinfo {author} {\bibfnamefont {E.}~\bibnamefont
  {Hiyama}}, \bibinfo {author} {\bibfnamefont {Y.}~\bibnamefont {Kino}}, \ and\
  \bibinfo {author} {\bibfnamefont {M.}~\bibnamefont {Kamimura}},\ }\href
  {\doibase 10.1016/S0146-6410(03)90015-9} {\bibfield  {journal} {\bibinfo
  {journal} {Prog. Part. Nucl. Phys.}\ }\textbf {\bibinfo {volume} {51}},\
  \bibinfo {pages} {223} (\bibinfo {year} {2003})}\BibitemShut {NoStop}%
\bibitem [{\citenamefont {Martinez~Torres}\ \emph
  {et~al.}(2008{\natexlab{b}})\citenamefont {Martinez~Torres}, \citenamefont
  {Khemchandani},\ and\ \citenamefont {Oset}}]{MartinezTorres:2007sr}%
  \BibitemOpen
  \bibfield  {author} {\bibinfo {author} {\bibfnamefont {A.}~\bibnamefont
  {Martinez~Torres}}, \bibinfo {author} {\bibfnamefont {K.~P.}\ \bibnamefont
  {Khemchandani}}, \ and\ \bibinfo {author} {\bibfnamefont {E.}~\bibnamefont
  {Oset}},\ }\href {\doibase 10.1103/PhysRevC.77.042203} {\bibfield  {journal}
  {\bibinfo  {journal} {Phys. Rev. C}\ }\textbf {\bibinfo {volume} {77}},\
  \bibinfo {pages} {042203} (\bibinfo {year} {2008}{\natexlab{b}})},\ \Eprint
  {http://arxiv.org/abs/0706.2330} {arXiv:0706.2330 [nucl-th]} \BibitemShut
  {NoStop}%
\bibitem [{\citenamefont {Martinez~Torres}\ \emph
  {et~al.}(2009{\natexlab{b}})\citenamefont {Martinez~Torres}, \citenamefont
  {Khemchandani}, \citenamefont {Meissner},\ and\ \citenamefont
  {Oset}}]{MartinezTorres:2009cw}%
  \BibitemOpen
  \bibfield  {author} {\bibinfo {author} {\bibfnamefont {A.}~\bibnamefont
  {Martinez~Torres}}, \bibinfo {author} {\bibfnamefont {K.~P.}\ \bibnamefont
  {Khemchandani}}, \bibinfo {author} {\bibfnamefont {U.-G.}\ \bibnamefont
  {Meissner}}, \ and\ \bibinfo {author} {\bibfnamefont {E.}~\bibnamefont
  {Oset}},\ }\href {\doibase 10.1140/epja/i2009-10834-2} {\bibfield  {journal}
  {\bibinfo  {journal} {Eur. Phys. J. A}\ }\textbf {\bibinfo {volume} {41}},\
  \bibinfo {pages} {361} (\bibinfo {year} {2009}{\natexlab{b}})},\ \Eprint
  {http://arxiv.org/abs/0902.3633} {arXiv:0902.3633 [nucl-th]} \BibitemShut
  {NoStop}%
\bibitem [{\citenamefont {Khemchandani}\ \emph {et~al.}(2009)\citenamefont
  {Khemchandani}, \citenamefont {Martinez~Torres},\ and\ \citenamefont
  {Oset}}]{Khemchandani:2009aj}%
  \BibitemOpen
  \bibfield  {author} {\bibinfo {author} {\bibfnamefont {K.~P.}\ \bibnamefont
  {Khemchandani}}, \bibinfo {author} {\bibfnamefont {A.}~\bibnamefont
  {Martinez~Torres}}, \ and\ \bibinfo {author} {\bibfnamefont {E.}~\bibnamefont
  {Oset}},\ }\href {\doibase 10.1016/j.physletb.2009.04.036} {\bibfield
  {journal} {\bibinfo  {journal} {Phys. Lett. B}\ }\textbf {\bibinfo {volume}
  {675}},\ \bibinfo {pages} {407} (\bibinfo {year} {2009})},\ \Eprint
  {http://arxiv.org/abs/0902.4425} {arXiv:0902.4425 [nucl-th]} \BibitemShut
  {NoStop}%
\bibitem [{\citenamefont {Chand}\ and\ \citenamefont
  {Dalitz}(1962)}]{Chand:1962ec}%
  \BibitemOpen
  \bibfield  {author} {\bibinfo {author} {\bibfnamefont {R.}~\bibnamefont
  {Chand}}\ and\ \bibinfo {author} {\bibfnamefont {R.~H.}\ \bibnamefont
  {Dalitz}},\ }\href {\doibase 10.1016/0003-4916(62)90113-6} {\bibfield
  {journal} {\bibinfo  {journal} {Annals Phys.}\ }\textbf {\bibinfo {volume}
  {20}},\ \bibinfo {pages} {1} (\bibinfo {year} {1962})}\BibitemShut {NoStop}%
\bibitem [{\citenamefont {Barrett}\ and\ \citenamefont
  {Deloff}(1999)}]{Barrett:1999cw}%
  \BibitemOpen
  \bibfield  {author} {\bibinfo {author} {\bibfnamefont {R.~C.}\ \bibnamefont
  {Barrett}}\ and\ \bibinfo {author} {\bibfnamefont {A.}~\bibnamefont
  {Deloff}},\ }\href {\doibase 10.1103/PhysRevC.60.025201} {\bibfield
  {journal} {\bibinfo  {journal} {Phys. Rev. C}\ }\textbf {\bibinfo {volume}
  {60}},\ \bibinfo {pages} {025201} (\bibinfo {year} {1999})}\BibitemShut
  {NoStop}%
\bibitem [{\citenamefont {Deloff}(2000)}]{Deloff:1999gc}%
  \BibitemOpen
  \bibfield  {author} {\bibinfo {author} {\bibfnamefont {A.}~\bibnamefont
  {Deloff}},\ }\href {\doibase 10.1103/PhysRevC.61.024004} {\bibfield
  {journal} {\bibinfo  {journal} {Phys. Rev. C}\ }\textbf {\bibinfo {volume}
  {61}},\ \bibinfo {pages} {024004} (\bibinfo {year} {2000})}\BibitemShut
  {NoStop}%
\bibitem [{\citenamefont {Kamalov}\ \emph {et~al.}(2001)\citenamefont
  {Kamalov}, \citenamefont {Oset},\ and\ \citenamefont
  {Ramos}}]{Kamalov:2000iy}%
  \BibitemOpen
  \bibfield  {author} {\bibinfo {author} {\bibfnamefont {S.~S.}\ \bibnamefont
  {Kamalov}}, \bibinfo {author} {\bibfnamefont {E.}~\bibnamefont {Oset}}, \
  and\ \bibinfo {author} {\bibfnamefont {A.}~\bibnamefont {Ramos}},\ }\href
  {\doibase 10.1016/S0375-9474(00)00709-0} {\bibfield  {journal} {\bibinfo
  {journal} {Nucl. Phys. A}\ }\textbf {\bibinfo {volume} {690}},\ \bibinfo
  {pages} {494} (\bibinfo {year} {2001})},\ \Eprint
  {http://arxiv.org/abs/nucl-th/0010054} {arXiv:nucl-th/0010054} \BibitemShut
  {NoStop}%
\bibitem [{\citenamefont {Roca}\ and\ \citenamefont
  {Oset}(2010)}]{Roca:2010tf}%
  \BibitemOpen
  \bibfield  {author} {\bibinfo {author} {\bibfnamefont {L.}~\bibnamefont
  {Roca}}\ and\ \bibinfo {author} {\bibfnamefont {E.}~\bibnamefont {Oset}},\
  }\href {\doibase 10.1103/PhysRevD.82.054013} {\bibfield  {journal} {\bibinfo
  {journal} {Phys. Rev. D}\ }\textbf {\bibinfo {volume} {82}},\ \bibinfo
  {pages} {054013} (\bibinfo {year} {2010})},\ \Eprint
  {http://arxiv.org/abs/1005.0283} {arXiv:1005.0283 [hep-ph]} \BibitemShut
  {NoStop}%
\bibitem [{\citenamefont {Yamagata-Sekihara}\ \emph {et~al.}(2010)\citenamefont
  {Yamagata-Sekihara}, \citenamefont {Roca},\ and\ \citenamefont
  {Oset}}]{Yamagata-Sekihara:2010muv}%
  \BibitemOpen
  \bibfield  {author} {\bibinfo {author} {\bibfnamefont {J.}~\bibnamefont
  {Yamagata-Sekihara}}, \bibinfo {author} {\bibfnamefont {L.}~\bibnamefont
  {Roca}}, \ and\ \bibinfo {author} {\bibfnamefont {E.}~\bibnamefont {Oset}},\
  }\href {\doibase 10.1103/PhysRevD.82.094017} {\bibfield  {journal} {\bibinfo
  {journal} {Phys. Rev. D}\ }\textbf {\bibinfo {volume} {82}},\ \bibinfo
  {pages} {094017} (\bibinfo {year} {2010})},\ \bibinfo {note} {[Erratum:
  Phys.Rev.D 85, 119905 (2012)]},\ \Eprint {http://arxiv.org/abs/1010.0525}
  {arXiv:1010.0525 [hep-ph]} \BibitemShut {NoStop}%
\bibitem [{\citenamefont {Xie}\ \emph {et~al.}(2011{\natexlab{a}})\citenamefont
  {Xie}, \citenamefont {Martinez~Torres}, \citenamefont {Oset},\ and\
  \citenamefont {Gonzalez}}]{Xie:2011uw}%
  \BibitemOpen
  \bibfield  {author} {\bibinfo {author} {\bibfnamefont {J.-J.}\ \bibnamefont
  {Xie}}, \bibinfo {author} {\bibfnamefont {A.}~\bibnamefont
  {Martinez~Torres}}, \bibinfo {author} {\bibfnamefont {E.}~\bibnamefont
  {Oset}}, \ and\ \bibinfo {author} {\bibfnamefont {P.}~\bibnamefont
  {Gonzalez}},\ }\href {\doibase 10.1103/PhysRevC.83.055204} {\bibfield
  {journal} {\bibinfo  {journal} {Phys. Rev. C}\ }\textbf {\bibinfo {volume}
  {83}},\ \bibinfo {pages} {055204} (\bibinfo {year} {2011}{\natexlab{a}})},\
  \Eprint {http://arxiv.org/abs/1101.1722} {arXiv:1101.1722 [nucl-th]}
  \BibitemShut {NoStop}%
\bibitem [{\citenamefont {Shen}\ and\ \citenamefont
  {Xie}(2023)}]{Shen:2022etd}%
  \BibitemOpen
  \bibfield  {author} {\bibinfo {author} {\bibfnamefont {Q.-H.}\ \bibnamefont
  {Shen}}\ and\ \bibinfo {author} {\bibfnamefont {J.-J.}\ \bibnamefont {Xie}},\
  }\href {\doibase 10.1103/PhysRevD.107.034019} {\bibfield  {journal} {\bibinfo
   {journal} {Phys. Rev. D}\ }\textbf {\bibinfo {volume} {107}},\ \bibinfo
  {pages} {034019} (\bibinfo {year} {2023})},\ \Eprint
  {http://arxiv.org/abs/2211.04911} {arXiv:2211.04911 [hep-ph]} \BibitemShut
  {NoStop}%
\bibitem [{\citenamefont {Workman}\ \emph {et~al.}(2022)\citenamefont {Workman}
  \emph {et~al.}}]{ParticleDataGroup:2022pth}%
  \BibitemOpen
  \bibfield  {author} {\bibinfo {author} {\bibfnamefont {R.~L.}\ \bibnamefont
  {Workman}} \emph {et~al.} (\bibinfo {collaboration} {Particle Data Group}),\
  }\href {\doibase 10.1093/ptep/ptac097} {\bibfield  {journal} {\bibinfo
  {journal} {PTEP}\ }\textbf {\bibinfo {volume} {2022}},\ \bibinfo {pages}
  {083C01} (\bibinfo {year} {2022})}\BibitemShut {NoStop}%
\bibitem [{\citenamefont {Dai}\ and\ \citenamefont {Oset}(2013)}]{Dai:2013uua}%
  \BibitemOpen
  \bibfield  {author} {\bibinfo {author} {\bibfnamefont {L.}~\bibnamefont
  {Dai}}\ and\ \bibinfo {author} {\bibfnamefont {E.}~\bibnamefont {Oset}},\
  }\href {\doibase 10.1140/epja/i2013-13130-8} {\bibfield  {journal} {\bibinfo
  {journal} {Eur. Phys. J. A}\ }\textbf {\bibinfo {volume} {49}},\ \bibinfo
  {pages} {130} (\bibinfo {year} {2013})},\ \Eprint
  {http://arxiv.org/abs/1306.2807} {arXiv:1306.2807 [hep-ph]} \BibitemShut
  {NoStop}%
\bibitem [{\citenamefont {Dai}\ \emph {et~al.}(2015)\citenamefont {Dai},
  \citenamefont {Xie},\ and\ \citenamefont {Oset}}]{Dai:2015cwa}%
  \BibitemOpen
  \bibfield  {author} {\bibinfo {author} {\bibfnamefont {L.-R.}\ \bibnamefont
  {Dai}}, \bibinfo {author} {\bibfnamefont {J.-J.}\ \bibnamefont {Xie}}, \ and\
  \bibinfo {author} {\bibfnamefont {E.}~\bibnamefont {Oset}},\ }\href {\doibase
  10.1103/PhysRevD.91.094013} {\bibfield  {journal} {\bibinfo  {journal} {Phys.
  Rev. D}\ }\textbf {\bibinfo {volume} {91}},\ \bibinfo {pages} {094013}
  (\bibinfo {year} {2015})},\ \Eprint {http://arxiv.org/abs/1503.02463}
  {arXiv:1503.02463 [hep-ph]} \BibitemShut {NoStop}%
\bibitem [{\citenamefont {Xie}\ \emph {et~al.}(2016{\natexlab{a}})\citenamefont
  {Xie}, \citenamefont {Oset},\ and\ \citenamefont {Geng}}]{Xie:2015isa}%
  \BibitemOpen
  \bibfield  {author} {\bibinfo {author} {\bibfnamefont {J.-J.}\ \bibnamefont
  {Xie}}, \bibinfo {author} {\bibfnamefont {E.}~\bibnamefont {Oset}}, \ and\
  \bibinfo {author} {\bibfnamefont {L.-S.}\ \bibnamefont {Geng}},\ }\href
  {\doibase 10.1103/PhysRevC.93.025202} {\bibfield  {journal} {\bibinfo
  {journal} {Phys. Rev. C}\ }\textbf {\bibinfo {volume} {93}},\ \bibinfo
  {pages} {025202} (\bibinfo {year} {2016}{\natexlab{a}})},\ \Eprint
  {http://arxiv.org/abs/1509.06469} {arXiv:1509.06469 [nucl-th]} \BibitemShut
  {NoStop}%
\bibitem [{\citenamefont {Xie}\ \emph {et~al.}(2016{\natexlab{b}})\citenamefont
  {Xie}, \citenamefont {Geng},\ and\ \citenamefont {Oset}}]{Xie:2016ghe}%
  \BibitemOpen
  \bibfield  {author} {\bibinfo {author} {\bibfnamefont {J.-J.}\ \bibnamefont
  {Xie}}, \bibinfo {author} {\bibfnamefont {L.-S.}\ \bibnamefont {Geng}}, \
  and\ \bibinfo {author} {\bibfnamefont {E.}~\bibnamefont {Oset}},\ }\href
  {\doibase 10.1051/epjconf/201613005021} {\bibfield  {journal} {\bibinfo
  {journal} {EPJ Web Conf.}\ }\textbf {\bibinfo {volume} {130}},\ \bibinfo
  {pages} {05021} (\bibinfo {year} {2016}{\natexlab{b}})}\BibitemShut {NoStop}%
\bibitem [{\citenamefont {Oset}\ \emph {et~al.}(2023)\citenamefont {Oset},
  \citenamefont {Dai},\ and\ \citenamefont {Geng}}]{Oset:2023hyt}%
  \BibitemOpen
  \bibfield  {author} {\bibinfo {author} {\bibfnamefont {E.}~\bibnamefont
  {Oset}}, \bibinfo {author} {\bibfnamefont {L.~R.}\ \bibnamefont {Dai}}, \
  and\ \bibinfo {author} {\bibfnamefont {L.~S.}\ \bibnamefont {Geng}},\
  }\href@noop {} {\  (\bibinfo {year} {2023})},\ \Eprint
  {http://arxiv.org/abs/2301.08532} {arXiv:2301.08532 [hep-ph]} \BibitemShut
  {NoStop}%
\bibitem [{\citenamefont {Lutz}\ and\ \citenamefont
  {Kolomeitsev}(2004)}]{Lutz:2003fm}%
  \BibitemOpen
  \bibfield  {author} {\bibinfo {author} {\bibfnamefont {M.~F.~M.}\
  \bibnamefont {Lutz}}\ and\ \bibinfo {author} {\bibfnamefont {E.~E.}\
  \bibnamefont {Kolomeitsev}},\ }\href {\doibase
  10.1016/j.nuclphysa.2003.11.009} {\bibfield  {journal} {\bibinfo  {journal}
  {Nucl. Phys. A}\ }\textbf {\bibinfo {volume} {730}},\ \bibinfo {pages} {392}
  (\bibinfo {year} {2004})},\ \Eprint {http://arxiv.org/abs/nucl-th/0307039}
  {arXiv:nucl-th/0307039} \BibitemShut {NoStop}%
\bibitem [{\citenamefont {Roca}\ \emph {et~al.}(2005)\citenamefont {Roca},
  \citenamefont {Oset},\ and\ \citenamefont {Singh}}]{Roca:2005nm}%
  \BibitemOpen
  \bibfield  {author} {\bibinfo {author} {\bibfnamefont {L.}~\bibnamefont
  {Roca}}, \bibinfo {author} {\bibfnamefont {E.}~\bibnamefont {Oset}}, \ and\
  \bibinfo {author} {\bibfnamefont {J.}~\bibnamefont {Singh}},\ }\href
  {\doibase 10.1103/PhysRevD.72.014002} {\bibfield  {journal} {\bibinfo
  {journal} {Phys. Rev. D}\ }\textbf {\bibinfo {volume} {72}},\ \bibinfo
  {pages} {014002} (\bibinfo {year} {2005})},\ \Eprint
  {http://arxiv.org/abs/hep-ph/0503273} {arXiv:hep-ph/0503273} \BibitemShut
  {NoStop}%
\bibitem [{\citenamefont {Ren}\ \emph {et~al.}(2018)\citenamefont {Ren},
  \citenamefont {Malabarba}, \citenamefont {Geng}, \citenamefont
  {Khemchandani},\ and\ \citenamefont {Mart\'\i{}nez~Torres}}]{Ren:2018pcd}%
  \BibitemOpen
  \bibfield  {author} {\bibinfo {author} {\bibfnamefont {X.-L.}\ \bibnamefont
  {Ren}}, \bibinfo {author} {\bibfnamefont {B.~B.}\ \bibnamefont {Malabarba}},
  \bibinfo {author} {\bibfnamefont {L.-S.}\ \bibnamefont {Geng}}, \bibinfo
  {author} {\bibfnamefont {K.~P.}\ \bibnamefont {Khemchandani}}, \ and\
  \bibinfo {author} {\bibfnamefont {A.}~\bibnamefont {Mart\'\i{}nez~Torres}},\
  }\href {\doibase 10.1016/j.physletb.2018.08.034} {\bibfield  {journal}
  {\bibinfo  {journal} {Phys. Lett. B}\ }\textbf {\bibinfo {volume} {785}},\
  \bibinfo {pages} {112} (\bibinfo {year} {2018})},\ \Eprint
  {http://arxiv.org/abs/1805.08330} {arXiv:1805.08330 [hep-ph]} \BibitemShut
  {NoStop}%
\bibitem [{\citenamefont {Dias}\ \emph {et~al.}(2017)\citenamefont {Dias},
  \citenamefont {Debastiani}, \citenamefont {Roca}, \citenamefont {Sakai},\
  and\ \citenamefont {Oset}}]{Dias:2017miz}%
  \BibitemOpen
  \bibfield  {author} {\bibinfo {author} {\bibfnamefont {J.~M.}\ \bibnamefont
  {Dias}}, \bibinfo {author} {\bibfnamefont {V.~R.}\ \bibnamefont
  {Debastiani}}, \bibinfo {author} {\bibfnamefont {L.}~\bibnamefont {Roca}},
  \bibinfo {author} {\bibfnamefont {S.}~\bibnamefont {Sakai}}, \ and\ \bibinfo
  {author} {\bibfnamefont {E.}~\bibnamefont {Oset}},\ }\href {\doibase
  10.1103/PhysRevD.96.094007} {\bibfield  {journal} {\bibinfo  {journal} {Phys.
  Rev. D}\ }\textbf {\bibinfo {volume} {96}},\ \bibinfo {pages} {094007}
  (\bibinfo {year} {2017})},\ \Eprint {http://arxiv.org/abs/1709.01372}
  {arXiv:1709.01372 [hep-ph]} \BibitemShut {NoStop}%
\bibitem [{\citenamefont {Bayar}\ \emph {et~al.}(2011)\citenamefont {Bayar},
  \citenamefont {Yamagata-Sekihara},\ and\ \citenamefont
  {Oset}}]{Bayar:2011qj}%
  \BibitemOpen
  \bibfield  {author} {\bibinfo {author} {\bibfnamefont {M.}~\bibnamefont
  {Bayar}}, \bibinfo {author} {\bibfnamefont {J.}~\bibnamefont
  {Yamagata-Sekihara}}, \ and\ \bibinfo {author} {\bibfnamefont
  {E.}~\bibnamefont {Oset}},\ }\href {\doibase 10.1103/PhysRevC.84.015209}
  {\bibfield  {journal} {\bibinfo  {journal} {Phys. Rev. C}\ }\textbf {\bibinfo
  {volume} {84}},\ \bibinfo {pages} {015209} (\bibinfo {year} {2011})},\
  \Eprint {http://arxiv.org/abs/1102.2854} {arXiv:1102.2854 [hep-ph]}
  \BibitemShut {NoStop}%
\bibitem [{\citenamefont {Xiao}\ \emph {et~al.}(2011)\citenamefont {Xiao},
  \citenamefont {Bayar},\ and\ \citenamefont {Oset}}]{Xiao:2011rc}%
  \BibitemOpen
  \bibfield  {author} {\bibinfo {author} {\bibfnamefont {C.~W.}\ \bibnamefont
  {Xiao}}, \bibinfo {author} {\bibfnamefont {M.}~\bibnamefont {Bayar}}, \ and\
  \bibinfo {author} {\bibfnamefont {E.}~\bibnamefont {Oset}},\ }\href {\doibase
  10.1103/PhysRevD.84.034037} {\bibfield  {journal} {\bibinfo  {journal} {Phys.
  Rev. D}\ }\textbf {\bibinfo {volume} {84}},\ \bibinfo {pages} {034037}
  (\bibinfo {year} {2011})},\ \Eprint {http://arxiv.org/abs/1106.0459}
  {arXiv:1106.0459 [hep-ph]} \BibitemShut {NoStop}%
\bibitem [{\citenamefont {Durkaya}\ and\ \citenamefont
  {Bayar}(2015)}]{Durkaya:2015wra}%
  \BibitemOpen
  \bibfield  {author} {\bibinfo {author} {\bibfnamefont {B.}~\bibnamefont
  {Durkaya}}\ and\ \bibinfo {author} {\bibfnamefont {M.}~\bibnamefont
  {Bayar}},\ }\href {\doibase 10.1103/PhysRevD.92.036006} {\bibfield  {journal}
  {\bibinfo  {journal} {Phys. Rev. D}\ }\textbf {\bibinfo {volume} {92}},\
  \bibinfo {pages} {036006} (\bibinfo {year} {2015})}\BibitemShut {NoStop}%
\bibitem [{\citenamefont {Martinez~Torres}\ \emph {et~al.}(2019)\citenamefont
  {Martinez~Torres}, \citenamefont {Khemchandani},\ and\ \citenamefont
  {Geng}}]{MartinezTorres:2018zbl}%
  \BibitemOpen
  \bibfield  {author} {\bibinfo {author} {\bibfnamefont {A.}~\bibnamefont
  {Martinez~Torres}}, \bibinfo {author} {\bibfnamefont {K.~P.}\ \bibnamefont
  {Khemchandani}}, \ and\ \bibinfo {author} {\bibfnamefont {L.-S.}\
  \bibnamefont {Geng}},\ }\href {\doibase 10.1103/PhysRevD.99.076017}
  {\bibfield  {journal} {\bibinfo  {journal} {Phys. Rev. D}\ }\textbf {\bibinfo
  {volume} {99}},\ \bibinfo {pages} {076017} (\bibinfo {year} {2019})},\
  \Eprint {http://arxiv.org/abs/1809.01059} {arXiv:1809.01059 [hep-ph]}
  \BibitemShut {NoStop}%
\bibitem [{\citenamefont {Sanchez~Sanchez}\ \emph {et~al.}(2018)\citenamefont
  {Sanchez~Sanchez}, \citenamefont {Geng}, \citenamefont {Lu}, \citenamefont
  {Hyodo},\ and\ \citenamefont {Valderrama}}]{SanchezSanchez:2017xtl}%
  \BibitemOpen
  \bibfield  {author} {\bibinfo {author} {\bibfnamefont {M.}~\bibnamefont
  {Sanchez~Sanchez}}, \bibinfo {author} {\bibfnamefont {L.-S.}\ \bibnamefont
  {Geng}}, \bibinfo {author} {\bibfnamefont {J.-X.}\ \bibnamefont {Lu}},
  \bibinfo {author} {\bibfnamefont {T.}~\bibnamefont {Hyodo}}, \ and\ \bibinfo
  {author} {\bibfnamefont {M.~P.}\ \bibnamefont {Valderrama}},\ }\href
  {\doibase 10.1103/PhysRevD.98.054001} {\bibfield  {journal} {\bibinfo
  {journal} {Phys. Rev. D}\ }\textbf {\bibinfo {volume} {98}},\ \bibinfo
  {pages} {054001} (\bibinfo {year} {2018})},\ \Eprint
  {http://arxiv.org/abs/1707.03802} {arXiv:1707.03802 [hep-ph]} \BibitemShut
  {NoStop}%
\bibitem [{\citenamefont {Xie}\ \emph {et~al.}(2011{\natexlab{b}})\citenamefont
  {Xie}, \citenamefont {Martinez~Torres},\ and\ \citenamefont
  {Oset}}]{Xie:2010ig}%
  \BibitemOpen
  \bibfield  {author} {\bibinfo {author} {\bibfnamefont {J.-J.}\ \bibnamefont
  {Xie}}, \bibinfo {author} {\bibfnamefont {A.}~\bibnamefont
  {Martinez~Torres}}, \ and\ \bibinfo {author} {\bibfnamefont {E.}~\bibnamefont
  {Oset}},\ }\href {\doibase 10.1103/PhysRevC.83.065207} {\bibfield  {journal}
  {\bibinfo  {journal} {Phys. Rev. C}\ }\textbf {\bibinfo {volume} {83}},\
  \bibinfo {pages} {065207} (\bibinfo {year} {2011}{\natexlab{b}})},\ \Eprint
  {http://arxiv.org/abs/1010.6164} {arXiv:1010.6164 [nucl-th]} \BibitemShut
  {NoStop}%
\bibitem [{\citenamefont {Martinez~Torres}\ \emph
  {et~al.}(2011{\natexlab{c}})\citenamefont {Martinez~Torres}, \citenamefont
  {Garzon}, \citenamefont {Oset},\ and\ \citenamefont
  {Dai}}]{MartinezTorres:2010ax}%
  \BibitemOpen
  \bibfield  {author} {\bibinfo {author} {\bibfnamefont {A.}~\bibnamefont
  {Martinez~Torres}}, \bibinfo {author} {\bibfnamefont {E.~J.}\ \bibnamefont
  {Garzon}}, \bibinfo {author} {\bibfnamefont {E.}~\bibnamefont {Oset}}, \ and\
  \bibinfo {author} {\bibfnamefont {L.~R.}\ \bibnamefont {Dai}},\ }\href
  {\doibase 10.1103/PhysRevD.83.116002} {\bibfield  {journal} {\bibinfo
  {journal} {Phys. Rev. D}\ }\textbf {\bibinfo {volume} {83}},\ \bibinfo
  {pages} {116002} (\bibinfo {year} {2011}{\natexlab{c}})},\ \Eprint
  {http://arxiv.org/abs/1012.2708} {arXiv:1012.2708 [hep-ph]} \BibitemShut
  {NoStop}%
\bibitem [{\citenamefont {Gamermann}\ \emph {et~al.}(2010)\citenamefont
  {Gamermann}, \citenamefont {Nieves}, \citenamefont {Oset},\ and\
  \citenamefont {Ruiz~Arriola}}]{Gamermann:2009uq}%
  \BibitemOpen
  \bibfield  {author} {\bibinfo {author} {\bibfnamefont {D.}~\bibnamefont
  {Gamermann}}, \bibinfo {author} {\bibfnamefont {J.}~\bibnamefont {Nieves}},
  \bibinfo {author} {\bibfnamefont {E.}~\bibnamefont {Oset}}, \ and\ \bibinfo
  {author} {\bibfnamefont {E.}~\bibnamefont {Ruiz~Arriola}},\ }\href {\doibase
  10.1103/PhysRevD.81.014029} {\bibfield  {journal} {\bibinfo  {journal} {Phys.
  Rev. D}\ }\textbf {\bibinfo {volume} {81}},\ \bibinfo {pages} {014029}
  (\bibinfo {year} {2010})},\ \Eprint {http://arxiv.org/abs/0911.4407}
  {arXiv:0911.4407 [hep-ph]} \BibitemShut {NoStop}%
\bibitem [{\citenamefont {Yamagata-Sekihara}\ \emph {et~al.}(2011)\citenamefont
  {Yamagata-Sekihara}, \citenamefont {Nieves},\ and\ \citenamefont
  {Oset}}]{Yamagata-Sekihara:2010kpd}%
  \BibitemOpen
  \bibfield  {author} {\bibinfo {author} {\bibfnamefont {J.}~\bibnamefont
  {Yamagata-Sekihara}}, \bibinfo {author} {\bibfnamefont {J.}~\bibnamefont
  {Nieves}}, \ and\ \bibinfo {author} {\bibfnamefont {E.}~\bibnamefont
  {Oset}},\ }\href {\doibase 10.1103/PhysRevD.83.014003} {\bibfield  {journal}
  {\bibinfo  {journal} {Phys. Rev. D}\ }\textbf {\bibinfo {volume} {83}},\
  \bibinfo {pages} {014003} (\bibinfo {year} {2011})},\ \Eprint
  {http://arxiv.org/abs/1007.3923} {arXiv:1007.3923 [hep-ph]} \BibitemShut
  {NoStop}%
\bibitem [{\citenamefont {Roca}(2011)}]{Roca:2011br}%
  \BibitemOpen
  \bibfield  {author} {\bibinfo {author} {\bibfnamefont {L.}~\bibnamefont
  {Roca}},\ }\href {\doibase 10.1103/PhysRevD.84.094006} {\bibfield  {journal}
  {\bibinfo  {journal} {Phys. Rev. D}\ }\textbf {\bibinfo {volume} {84}},\
  \bibinfo {pages} {094006} (\bibinfo {year} {2011})},\ \Eprint
  {http://arxiv.org/abs/1107.5489} {arXiv:1107.5489 [hep-ph]} \BibitemShut
  {NoStop}%
\bibitem [{\citenamefont {Geng}\ \emph {et~al.}(2007)\citenamefont {Geng},
  \citenamefont {Oset}, \citenamefont {Roca},\ and\ \citenamefont
  {Oller}}]{Geng:2006yb}%
  \BibitemOpen
  \bibfield  {author} {\bibinfo {author} {\bibfnamefont {L.~S.}\ \bibnamefont
  {Geng}}, \bibinfo {author} {\bibfnamefont {E.}~\bibnamefont {Oset}}, \bibinfo
  {author} {\bibfnamefont {L.}~\bibnamefont {Roca}}, \ and\ \bibinfo {author}
  {\bibfnamefont {J.~A.}\ \bibnamefont {Oller}},\ }\href {\doibase
  10.1103/PhysRevD.75.014017} {\bibfield  {journal} {\bibinfo  {journal} {Phys.
  Rev. D}\ }\textbf {\bibinfo {volume} {75}},\ \bibinfo {pages} {014017}
  (\bibinfo {year} {2007})},\ \Eprint {http://arxiv.org/abs/hep-ph/0610217}
  {arXiv:hep-ph/0610217} \BibitemShut {NoStop}%
\bibitem [{\citenamefont {Guo}\ \emph {et~al.}(2006)\citenamefont {Guo},
  \citenamefont {Ping}, \citenamefont {Shen}, \citenamefont {Chiang},\ and\
  \citenamefont {Zou}}]{Guo:2005wp}%
  \BibitemOpen
  \bibfield  {author} {\bibinfo {author} {\bibfnamefont {F.-K.}\ \bibnamefont
  {Guo}}, \bibinfo {author} {\bibfnamefont {R.-G.}\ \bibnamefont {Ping}},
  \bibinfo {author} {\bibfnamefont {P.-N.}\ \bibnamefont {Shen}}, \bibinfo
  {author} {\bibfnamefont {H.-C.}\ \bibnamefont {Chiang}}, \ and\ \bibinfo
  {author} {\bibfnamefont {B.-S.}\ \bibnamefont {Zou}},\ }\href {\doibase
  10.1016/j.nuclphysa.2006.04.008} {\bibfield  {journal} {\bibinfo  {journal}
  {Nucl. Phys. A}\ }\textbf {\bibinfo {volume} {773}},\ \bibinfo {pages} {78}
  (\bibinfo {year} {2006})},\ \Eprint {http://arxiv.org/abs/hep-ph/0509050}
  {arXiv:hep-ph/0509050} \BibitemShut {NoStop}%
\bibitem [{\citenamefont {Sun}\ \emph {et~al.}(2022)\citenamefont {Sun},
  \citenamefont {Fan},\ and\ \citenamefont {Cao}}]{Sun:2022cxf}%
  \BibitemOpen
  \bibfield  {author} {\bibinfo {author} {\bibfnamefont {B.-X.}\ \bibnamefont
  {Sun}}, \bibinfo {author} {\bibfnamefont {Y.-Y.}\ \bibnamefont {Fan}}, \ and\
  \bibinfo {author} {\bibfnamefont {Q.-Q.}\ \bibnamefont {Cao}},\ }\href@noop
  {} {\  (\bibinfo {year} {2022})},\ \Eprint {http://arxiv.org/abs/2206.02961}
  {arXiv:2206.02961 [hep-ph]} \BibitemShut {NoStop}%
\bibitem [{\citenamefont {Anisovich}\ \emph
  {et~al.}(2001{\natexlab{a}})\citenamefont {Anisovich}, \citenamefont {Baker},
  \citenamefont {Batty}, \citenamefont {Bugg}, \citenamefont {Nikonov},
  \citenamefont {Sarantsev}, \citenamefont {Sarantsev},\ and\ \citenamefont
  {Zou}}]{Anisovich:2001pn}%
  \BibitemOpen
  \bibfield  {author} {\bibinfo {author} {\bibfnamefont {A.~V.}\ \bibnamefont
  {Anisovich}}, \bibinfo {author} {\bibfnamefont {C.~A.}\ \bibnamefont
  {Baker}}, \bibinfo {author} {\bibfnamefont {C.~J.}\ \bibnamefont {Batty}},
  \bibinfo {author} {\bibfnamefont {D.~V.}\ \bibnamefont {Bugg}}, \bibinfo
  {author} {\bibfnamefont {V.~A.}\ \bibnamefont {Nikonov}}, \bibinfo {author}
  {\bibfnamefont {A.~V.}\ \bibnamefont {Sarantsev}}, \bibinfo {author}
  {\bibfnamefont {V.~V.}\ \bibnamefont {Sarantsev}}, \ and\ \bibinfo {author}
  {\bibfnamefont {B.~S.}\ \bibnamefont {Zou}},\ }\href {\doibase
  10.1016/S0370-2693(01)01017-6} {\bibfield  {journal} {\bibinfo  {journal}
  {Phys. Lett. B}\ }\textbf {\bibinfo {volume} {517}},\ \bibinfo {pages} {261}
  (\bibinfo {year} {2001}{\natexlab{a}})},\ \Eprint
  {http://arxiv.org/abs/1110.0278} {arXiv:1110.0278 [hep-ex]} \BibitemShut
  {NoStop}%
\bibitem [{\citenamefont {Anisovich}\ \emph {et~al.}(2000)\citenamefont
  {Anisovich} \emph {et~al.}}]{Anisovich:2000ut}%
  \BibitemOpen
  \bibfield  {author} {\bibinfo {author} {\bibfnamefont {A.~V.}\ \bibnamefont
  {Anisovich}} \emph {et~al.},\ }\href {\doibase 10.1016/S0370-2693(00)01018-2}
  {\bibfield  {journal} {\bibinfo  {journal} {Phys. Lett. B}\ }\textbf
  {\bibinfo {volume} {491}},\ \bibinfo {pages} {47} (\bibinfo {year} {2000})},\
  \Eprint {http://arxiv.org/abs/1109.0883} {arXiv:1109.0883 [hep-ex]}
  \BibitemShut {NoStop}%
\bibitem [{\citenamefont {Barberis}\ \emph {et~al.}(2000)\citenamefont
  {Barberis} \emph {et~al.}}]{WA102:1999ybu}%
  \BibitemOpen
  \bibfield  {author} {\bibinfo {author} {\bibfnamefont {D.}~\bibnamefont
  {Barberis}} \emph {et~al.} (\bibinfo {collaboration} {WA102}),\ }\href
  {\doibase 10.1016/S0370-2693(99)01394-5} {\bibfield  {journal} {\bibinfo
  {journal} {Phys. Lett. B}\ }\textbf {\bibinfo {volume} {471}},\ \bibinfo
  {pages} {435} (\bibinfo {year} {2000})},\ \Eprint
  {http://arxiv.org/abs/hep-ex/9911038} {arXiv:hep-ex/9911038} \BibitemShut
  {NoStop}%
\bibitem [{\citenamefont {Karch}\ \emph {et~al.}(1992)\citenamefont {Karch}
  \emph {et~al.}}]{CrystalBall:1991zkb}%
  \BibitemOpen
  \bibfield  {author} {\bibinfo {author} {\bibfnamefont {K.}~\bibnamefont
  {Karch}} \emph {et~al.} (\bibinfo {collaboration} {Crystal Ball}),\ }\href
  {\doibase 10.1007/BF01881706} {\bibfield  {journal} {\bibinfo  {journal} {Z.
  Phys. C}\ }\textbf {\bibinfo {volume} {54}},\ \bibinfo {pages} {33} (\bibinfo
  {year} {1992})}\BibitemShut {NoStop}%
\bibitem [{\citenamefont {Isgur}\ and\ \citenamefont
  {Paton}(1985)}]{Isgur:1984bm}%
  \BibitemOpen
  \bibfield  {author} {\bibinfo {author} {\bibfnamefont {N.}~\bibnamefont
  {Isgur}}\ and\ \bibinfo {author} {\bibfnamefont {J.~E.}\ \bibnamefont
  {Paton}},\ }\href {\doibase 10.1103/PhysRevD.31.2910} {\bibfield  {journal}
  {\bibinfo  {journal} {Phys. Rev. D}\ }\textbf {\bibinfo {volume} {31}},\
  \bibinfo {pages} {2910} (\bibinfo {year} {1985})}\BibitemShut {NoStop}%
\bibitem [{\citenamefont {Godfrey}\ and\ \citenamefont
  {Isgur}(1985)}]{Godfrey:1985xj}%
  \BibitemOpen
  \bibfield  {author} {\bibinfo {author} {\bibfnamefont {S.}~\bibnamefont
  {Godfrey}}\ and\ \bibinfo {author} {\bibfnamefont {N.}~\bibnamefont
  {Isgur}},\ }\href {\doibase 10.1103/PhysRevD.32.189} {\bibfield  {journal}
  {\bibinfo  {journal} {Phys. Rev. D}\ }\textbf {\bibinfo {volume} {32}},\
  \bibinfo {pages} {189} (\bibinfo {year} {1985})}\BibitemShut {NoStop}%
\bibitem [{\citenamefont {Vijande}\ \emph {et~al.}(2005)\citenamefont
  {Vijande}, \citenamefont {Fernandez},\ and\ \citenamefont
  {Valcarce}}]{Vijande:2004he}%
  \BibitemOpen
  \bibfield  {author} {\bibinfo {author} {\bibfnamefont {J.}~\bibnamefont
  {Vijande}}, \bibinfo {author} {\bibfnamefont {F.}~\bibnamefont {Fernandez}},
  \ and\ \bibinfo {author} {\bibfnamefont {A.}~\bibnamefont {Valcarce}},\
  }\href {\doibase 10.1088/0954-3899/31/5/017} {\bibfield  {journal} {\bibinfo
  {journal} {J. Phys. G}\ }\textbf {\bibinfo {volume} {31}},\ \bibinfo {pages}
  {481} (\bibinfo {year} {2005})},\ \Eprint
  {http://arxiv.org/abs/hep-ph/0411299} {arXiv:hep-ph/0411299} \BibitemShut
  {NoStop}%
\bibitem [{\citenamefont {Wang}\ \emph {et~al.}(2015)\citenamefont {Wang},
  \citenamefont {Pang}, \citenamefont {Liu},\ and\ \citenamefont
  {Matsuki}}]{Wang:2014sea}%
  \BibitemOpen
  \bibfield  {author} {\bibinfo {author} {\bibfnamefont {B.}~\bibnamefont
  {Wang}}, \bibinfo {author} {\bibfnamefont {C.-Q.}\ \bibnamefont {Pang}},
  \bibinfo {author} {\bibfnamefont {X.}~\bibnamefont {Liu}}, \ and\ \bibinfo
  {author} {\bibfnamefont {T.}~\bibnamefont {Matsuki}},\ }\href {\doibase
  10.1103/PhysRevD.91.014025} {\bibfield  {journal} {\bibinfo  {journal} {Phys.
  Rev. D}\ }\textbf {\bibinfo {volume} {91}},\ \bibinfo {pages} {014025}
  (\bibinfo {year} {2015})},\ \Eprint {http://arxiv.org/abs/1410.3930}
  {arXiv:1410.3930 [hep-ph]} \BibitemShut {NoStop}%
\bibitem [{\citenamefont {Li}\ and\ \citenamefont {Wang}(2009)}]{Li:2009rka}%
  \BibitemOpen
  \bibfield  {author} {\bibinfo {author} {\bibfnamefont {D.-M.}\ \bibnamefont
  {Li}}\ and\ \bibinfo {author} {\bibfnamefont {E.}~\bibnamefont {Wang}},\
  }\href {\doibase 10.1140/epjc/s10052-009-1106-z} {\bibfield  {journal}
  {\bibinfo  {journal} {Eur. Phys. J. C}\ }\textbf {\bibinfo {volume} {63}},\
  \bibinfo {pages} {297} (\bibinfo {year} {2009})},\ \Eprint
  {http://arxiv.org/abs/0904.1252} {arXiv:0904.1252 [hep-ph]} \BibitemShut
  {NoStop}%
\bibitem [{\citenamefont {Anisovich}\ \emph
  {et~al.}(2001{\natexlab{b}})\citenamefont {Anisovich}, \citenamefont {Baker},
  \citenamefont {Batty}, \citenamefont {Bugg}, \citenamefont {Nikonov},
  \citenamefont {Sarantsev}, \citenamefont {Sarantsev},\ and\ \citenamefont
  {Zou}}]{Anisovich:2001hj}%
  \BibitemOpen
  \bibfield  {author} {\bibinfo {author} {\bibfnamefont {A.~V.}\ \bibnamefont
  {Anisovich}}, \bibinfo {author} {\bibfnamefont {C.~A.}\ \bibnamefont
  {Baker}}, \bibinfo {author} {\bibfnamefont {C.~J.}\ \bibnamefont {Batty}},
  \bibinfo {author} {\bibfnamefont {D.~V.}\ \bibnamefont {Bugg}}, \bibinfo
  {author} {\bibfnamefont {V.~A.}\ \bibnamefont {Nikonov}}, \bibinfo {author}
  {\bibfnamefont {A.~V.}\ \bibnamefont {Sarantsev}}, \bibinfo {author}
  {\bibfnamefont {V.~V.}\ \bibnamefont {Sarantsev}}, \ and\ \bibinfo {author}
  {\bibfnamefont {B.~S.}\ \bibnamefont {Zou}},\ }\href {\doibase
  10.1016/S0370-2693(01)00089-2} {\bibfield  {journal} {\bibinfo  {journal}
  {Phys. Lett. B}\ }\textbf {\bibinfo {volume} {500}},\ \bibinfo {pages} {222}
  (\bibinfo {year} {2001}{\natexlab{b}})},\ \Eprint
  {http://arxiv.org/abs/1109.6433} {arXiv:1109.6433 [hep-ex]} \BibitemShut
  {NoStop}%
\bibitem [{\citenamefont {Barnes}\ \emph {et~al.}(1995)\citenamefont {Barnes},
  \citenamefont {Close},\ and\ \citenamefont {Swanson}}]{Barnes:1995hc}%
  \BibitemOpen
  \bibfield  {author} {\bibinfo {author} {\bibfnamefont {T.}~\bibnamefont
  {Barnes}}, \bibinfo {author} {\bibfnamefont {F.~E.}\ \bibnamefont {Close}}, \
  and\ \bibinfo {author} {\bibfnamefont {E.~S.}\ \bibnamefont {Swanson}},\
  }\href {\doibase 10.1103/PhysRevD.52.5242} {\bibfield  {journal} {\bibinfo
  {journal} {Phys. Rev. D}\ }\textbf {\bibinfo {volume} {52}},\ \bibinfo
  {pages} {5242} (\bibinfo {year} {1995})},\ \Eprint
  {http://arxiv.org/abs/hep-ph/9501405} {arXiv:hep-ph/9501405} \BibitemShut
  {NoStop}%
\end{thebibliography}%

\end{document}